\documentclass[%
 preprint,
 amsmath,amssymb,
 aip
]{revtex4-2}

\usepackage{graphicx}
\usepackage{dcolumn}
\usepackage{wrapfig}
\usepackage{bm}
\usepackage{multirow}
\usepackage{color} 

\begin{document}

\title{Control of electron beam current, charge and energy spread using density downramp injection in laser wakefield accelerators}

\author{C\'eline Hue, Yang Wan, Eitan Levine, and Victor Malka}

\affiliation{Department of Physics of Complex Systems, Weizmann Institute of Science, Rehovot, Israel
}%

\date{\today}

\begin{abstract}
Density dowmramp injection has been demonstrated to be an elegant and efficient approach for generating high quality electron beams in laser wakefield accelerators. Yet, the charge of the produced beam is tens of pC per Joule of laser energy, still limiting its use for a wider range of applications. The possibility of generating high charge beam while keeping a good beam quality, stays to be explored. Moreover, despite previous studies focused on separate physical processes such as beam loading which affects the uniformity of the acceleration field and thus the energy spread of the trapped electrons, repulsive force from the rear spike of the bubble which reduces the transverse momentum $p_\perp$ of the trapped electrons and results in small beam emmittance, and the laser evolution when travelling in plasma. A more general investigation of the plasma density parameters on the final beam properties is required. In this work, we demonstrate that the current profile of the injected electron beam is directly correlated with the density transition parameters, which further affects the beam charge and energy spread. By fine-tuning the plasma density parameters, high-charge (up to several hundreds of pC) and low-energy-spread (around 1\% FWHM) electron beams can be obtained. All these results are supported by large-scale three-dimensional particle-in-cell simulations. 
\end{abstract}

\maketitle

\section{\label{sec:Intro}Introduction}
Laser wakefield acceleration (LWFA) is one of the most promising accelerator technologies which offers acceleration gradient more than three orders of magnitude higher than conventional accelerators \cite{Faure2004,Geddes2004,Mangles2004}. In LWFAs, plasma electrons are pushed outwards by the ponderomotive force of the intense laser pulse forming a plasma wave \cite{Tajima1979}, that travels together with the laser pulse at relativistic speeds, and therefore suitable for compact and efficient acceleration of trapped electrons. Numerous injection methods have been proposed (see \cite{Malka2012} for a review article on injection method), explored and demonstrated to improve the beam quality. Over the past decades, electron injection using a sharp plasma density transition profile has been demonstrated to be propitious in the generation of high-quality electron beams in LWFAs \cite{Bulanov1998,Thaury2015,Buck2013ShockInjectionExp,Couperus2021Hybrid}, where plasma electrons are trapped due to the wave-breaking induced by the longitudinal expansion of the plasma wave structure when propagating along the density down-ramp. This approach can be easily implemented by inserting a sharp blade on top of the gas nozzle, and has been used recently to demonstrate free electron lasing with a LWFA high-quality electron beam of 10-50 pC \cite{Wang2021FEL}. However, more studies are still required to yet improve the laser-to-electrons conversion efficiency, always suitable for applications including for example  radiotherapy\cite{Yeboah_2002_250MeVEleTherapy, Glinnec2005, Malka2008,Fuchs_2009} while constraints on other parameters such as energy spread, and emittance can be relaxed. A more comprehensive study covering a broad range of input conditions stays yet very difficult because of the non-linearity of the interaction. Previously, this challenge has been performed with scans of given parameters \cite{Massimo_2018,Massimo_2017}. Yet parametric study over a wider range of scanned parameters is needed for a better understanding and optimisation of this important injection mechanism method.

In this article, an approach for tailoring the final beam parameters (charge $Q$, beam energy $E$, and energy spread $\delta E$) by tuning three parameters of the plasma density profile (the density ratio K, the downramp length L and the downramp position $\text{Pos}_d$) as shown in Figure 1, was studied by mean of numerical simulations performed with 3D PIC code FBPIC\cite{LEHE201666}, in a way that the physics behind it is being understood. 
\begin{figure}
\includegraphics[width=0.8\textwidth]{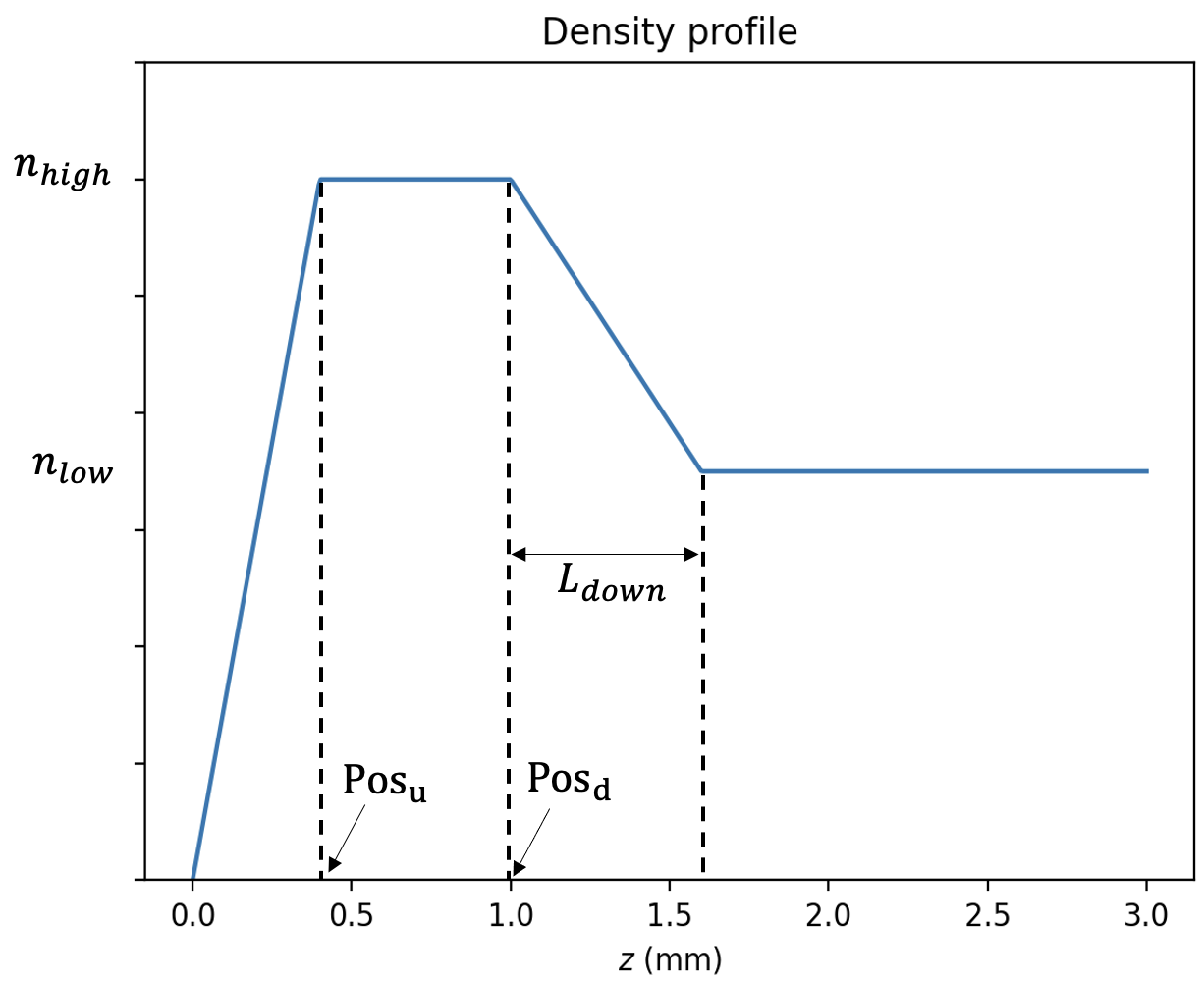}
\caption{\label{fig:density_exemple} The plasma density profile used in the study, where laser pulse propagates along the z positive direction.}
\end{figure}
 
The considered laser parameters are common with those delivered by typical  hundred TW - few tens of fs laser chain. Here we took those of the HIGGINS laser system at Weizmann Institute of Science \cite{Kroupp2022}, with a pulse duration of 30 fs, a focal spot waist of 18 $\mu$m and a normalized vector potential $a_0$ of 2.2, corresponding to an on-target laser energy of 1.6 J. The simulations were performed in 3D cylindrical grid in a cylindrical geometry with number of azimuthal mode as $2$. We choose a mesh resolution $\Delta z = 0.30c/\omega_0$ and $\Delta r = 0.73 c/\omega_0$ in the longitudinal and radial direction, respectively, with integration time step $\Delta t = 0.22\omega_0^{-1}$ where $\omega_0=2\pi c/\lambda$ is the laser central frequency. The results shown in the following have been obtained with $45$ particles per mesh cell. In the simulations, the plasma was set to be pre-ionized which allowed us to study solely the impact of the plasma profile on the injected beam parameters. Despite further precision, the $n_{low}$ of plasma is fixed to be $2.5\times 10^{18}\text{cm}^{\text{-3}}$.

\begin{figure*}[ht!]
\includegraphics[width=1.0
\textwidth]{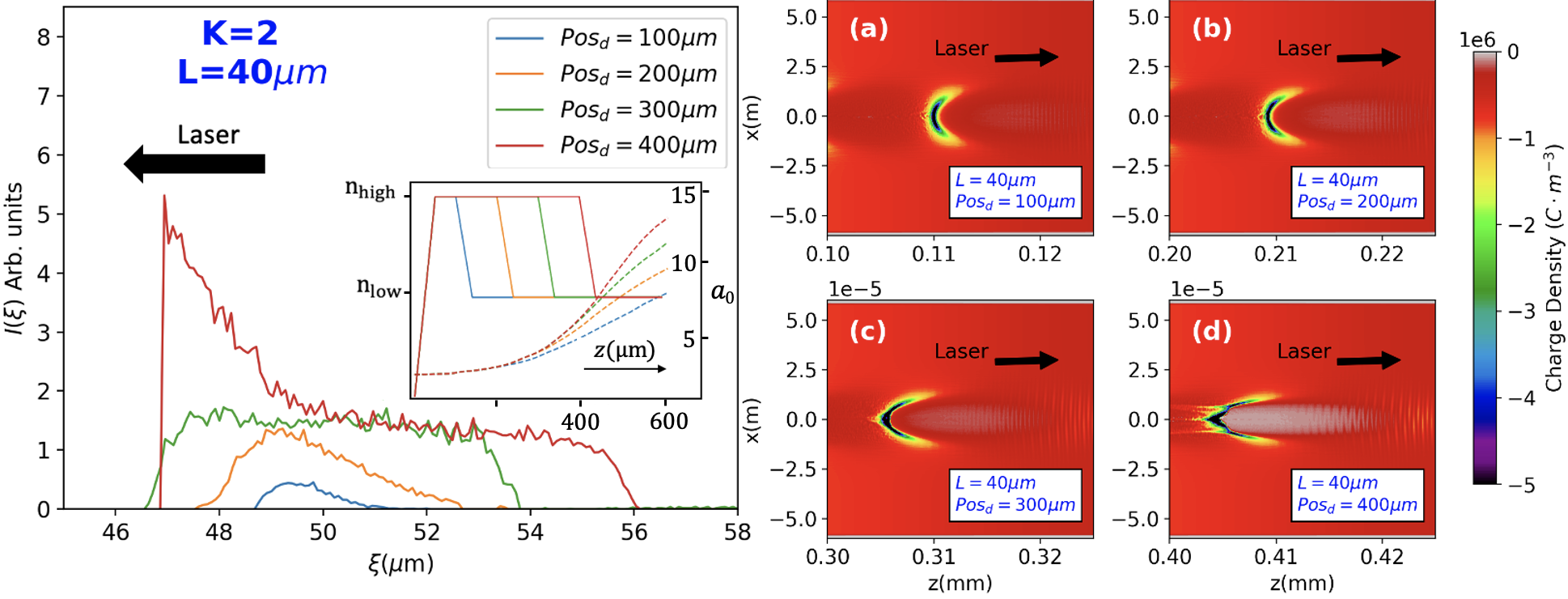}
\caption{\label{fig: I(xi) 40} Left side: beam current distribution at the plasma exit for  ratio $K = n_\text{high}/n_e$ equals to $2$ and downramp length $L$ equals to $40\mu m$. The plasma density and laser $a_0$ evolution are illustrated in the inset figure. Left side: Plasma charge density of the wakefield structures taken at positions where the rear bubble part corresponds to the middle of the density downramp.}
\end{figure*}

In this study, it is found that the beam current distribution $I(\xi)=c\times q(\xi)$, is determined by the density downramp parameters, which further affects the injected beam charge and energy performance. Here, $\xi = v_d t -z \approx ct-z$, $v_d$ is the group velocity of the laser driver, $e$ the elementary charge, $c$ the speed of light and $q(\xi)$ the beam charge per unit length. The article is organized as follows: we first show in section \ref{sec: tailoring} how one can tailor the beam current distribution $I(\xi)$ by changing the plasma density profile. In section \ref{sec:I(z) and energy spread}, the dependence of beam energy evolution with the beam current $I(\xi)$ is discussed. This section also reports on achievement of few hundreds of MeV and pC electron beams with relatively good beam quality. Section IV explains the maximum limit of injected beam charge and its effect on the final beam energy and energy spread. Section V present conclusions and perspectives.

\section{\label{sec: tailoring}Tailoring the beam current distribution with different plasma profile}
In this section, we show how one can tailor the beam current distribution $I(\xi)$ by controlling the three parameters that described the plasma density profile: $K$, $L_\text{down}$ and $\text{Pos}_d$. These parameters can be easily tuned experimentally (for example by changing the blade or string-like obstacles\cite{Thaury2015,BUCK2013,Schmid2010} located on the path of the gas jet). 

It is found that the value of the laser intensity at the position of the density down-ramp and accordingly its corresponding wakefield strength play decisive roles in shaping $I(\xi)$. The four simulations reported on figure \ref{fig: I(xi) 40} show the electron beam current distributions at the exit of the plasma together with the normalized laser potential $a_0$ in the plasma density region of interest (both are reported in the inset). Only the donwramp initial position $\text{Pos}_d$ is changed in the simulations while keeping the downramp density ratio $K = 2$ and the donwramp length $L=40$ $\mu$m fixed.

The beam charge $Q$ increases with higher $a_0$ across the downramp region with different shape of beam current $I(\xi)$ changing from triangle-like (blue and orange lines) to rectangle-like (green lines), then to a piecewise function of two segments consist of a peak and a constant segment(red lines). The beam current distribution behaviors are mainly due to the wakefield strength in the plasma density downramp. The higher the laser intensity, the stronger the wakefield strength and the easier the injection (higher beam current $I(\xi)$ and charge Q). The right-sided sub-figures in fig.\ref{fig: I(xi) 40} shows the plasma charge density of the wakefield taken at positions at the middle of the density downramp. One can see that the bubble structure become clearer as $\text{Pos}_d$ increases. 

\begin{figure*}[ht!]
\includegraphics[width=0.9
\textwidth]{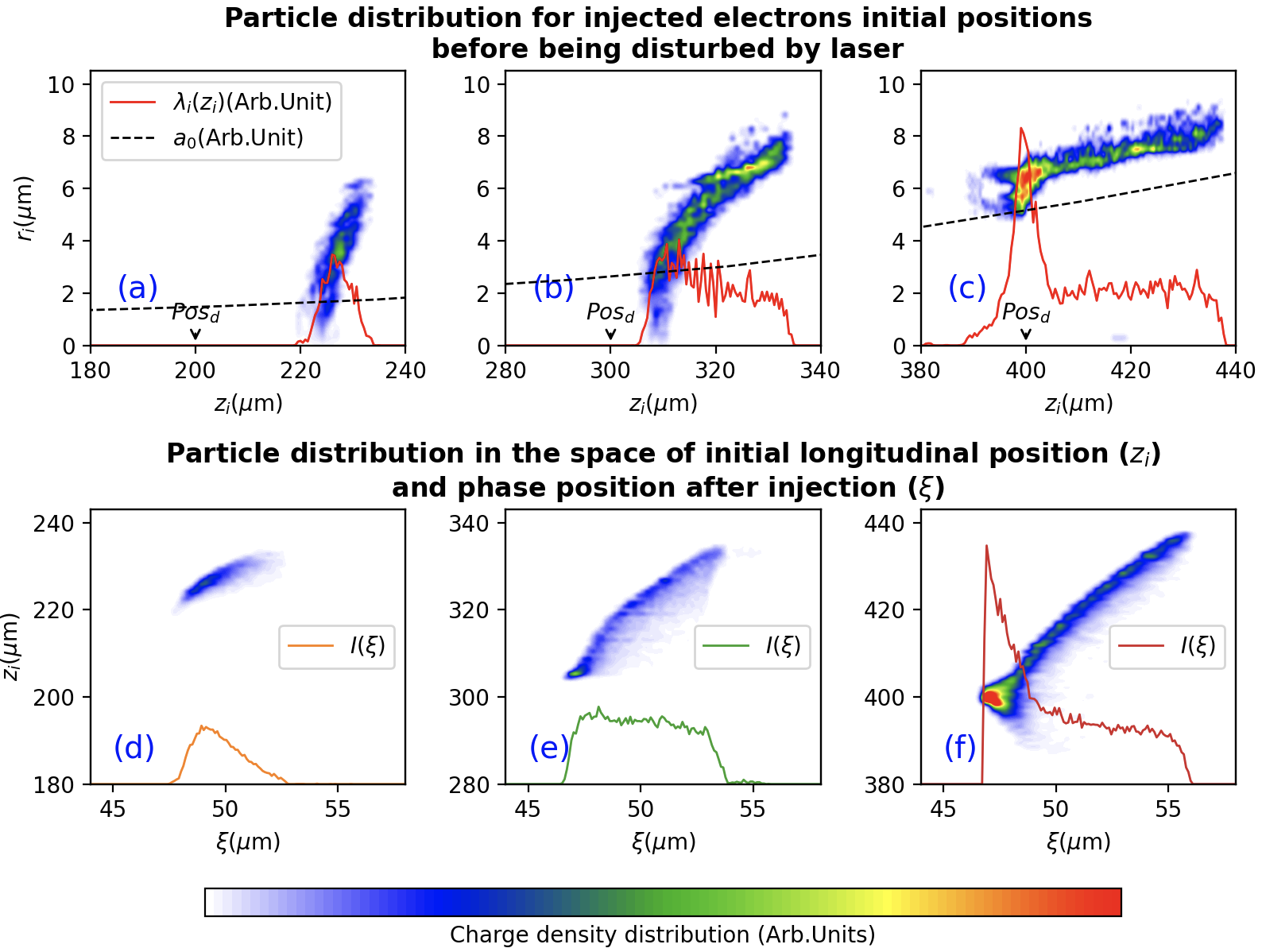}
\caption{\label{fig: Particle initial distribution} Particle distribution for injected particles initial positions before being disturbed by laser for each referred simulation in $z_i-r_i$ plane: sub-figures (a), (b) and (c), and particle distributions in the space constructed of their initial longitudinal position $z_i$, and the phase position in the wakefield after the injection $\xi$: sub-figures (d), (e) and (f).}
\end{figure*}

For a better understanding on how the laser and wakefield intensity influence the injection, we investigate the initial positions ($z_i$, $r_i$) of the injected electrons before being affected by the laser, together with the correlation between the initial longitudinal position $z_i$ and final phase positions $\xi$ after injection for three different simulations corresponding to the orange, green and red lines of Figure \ref{fig: I(xi) 40}. Electrons from the injected beam are traced back to their original positions in plasma and their distributions in $z_i-r_i$ plane are shown in Figure \ref{fig: Particle initial distribution}(a)-(c), where the downramp starts from $200\mu m$, $300\mu m$ and $400\mu m$, and ends at $240\mu m$, $340\mu m$ and $440\mu m$, respectively . The evolution of $a_0$ in these regions are plotted as dashed black lines showing an increase of $a_0$ for higher $\text{Pos}_d$. The line-outs of the charge density distributions (charge per unit length across $z_i$) over the longitudinal coordinate $z$, which we name here as $\lambda_i(z_i)$, are plotted as red solid lines.

For larger values of $\text{Pos}_d$ as shown in Fig.3(c), thanks to the nonlinear processes (such as relativistic self focusing \cite{Sum1987_selffocusing} or self steepening effects \cite{Vieira_2010} that increase with the propagation distance, the blow-out/bubble regime throughout the downramp region is authorized and the injected particles cover the whole downramp region (from $400\mu m$ to $440\mu m$ for the mentioned case). A small portion of particles before downramp region ($z_i<400\mu m$) is also involved in the injection process. The shape of particle initial distribution in $z_i-r_i$ plane (Figure \ref{fig: Particle initial distribution}.c) is similar to  what is explained in \cite{XU2017}, that when the blow-out is complete throughout downramp region, and the ramp decreases gradually ($l\equiv |n_p/(dn_p/dz)|\gg c/\omega_p$), only particles situated near a fixed transverse position $r_i\simeq\kappa r_m$, where $r_m$ is the maximum transverse radius of the bubble, are able to be injected. Similar to the \cite{XU2017}, $\kappa r_m$ slightly increases with the decrease of plasma density $n_e(z_i)$ in the downramp. Number of particle per longitudinal slice $\Delta z_i$ issued around $\kappa r_m$ is constant through the downramp thus $\lambda_i(z_i)$ line-out shows bloc-like constant profile. The charge density peak located at the beginning of downramp is probably due to the rapid changes of the the wakefield structure which cause extra particles be accelerated and injected. 

For smaller $\text{Pos}_d$ of 200 $\mu$m and 300 $\mu$m as shown in Fig.3(a) and (b) respectively, a complete blow-out is not reached at the downramp, and injection starts later in respect to the above case. Only particles initially located from middle to the end of the downramp are injected. Attentions are paid on $\lambda_i(z_i)$ line-outs for the two cases of interest. For $\text{Pos}_d=200 \mu m$, where the wakefield is still weak, injection occurs at a position closer to the rear of the downramp and $\lambda_i(z_i)$ has a triangle-like shape. For $\text{Pos}_d=300 \mu m$, the wakefield is close to the blow-out regime, and the $\lambda_i(z_i)$ shows to be a trapezoid shape. In these two cases, particles with smaller initial transverse positions (nearer to the axis) are also engaged in the injection process. These particles are accelerated and injected inside the wakefield without going along the sheath of bubbles in plasma wakes. 

We further investigate and report on the beam charge density distributions in the $\xi-z_i$ plane in Figures \ref{fig: Particle initial distribution}.d), \ref{fig: Particle initial distribution}.e) and \ref{fig: Particle initial distribution}.f), that correspond respectively to $\text{Pos}_d=200 \mu m$, $\text{Pos}_d=300 \mu m$ and $\text{Pos}_d=400 \mu m$, where the normalized beam current is also plotted. Almost along the whole bunch duration, a linear correlation is observed between the initial longitudinal $z_i$ and the injected phase position $\xi$. Such a linearity indicates that $\lambda_i (z_i )$ line-out over $z_i$ takes similar shape as the final beam current $I(\xi)$ as shown in the upper and lower sub-figures of the figure \ref{fig: Particle initial distribution}.

\begin{figure}[t]
\includegraphics[width=0.80
\textwidth]{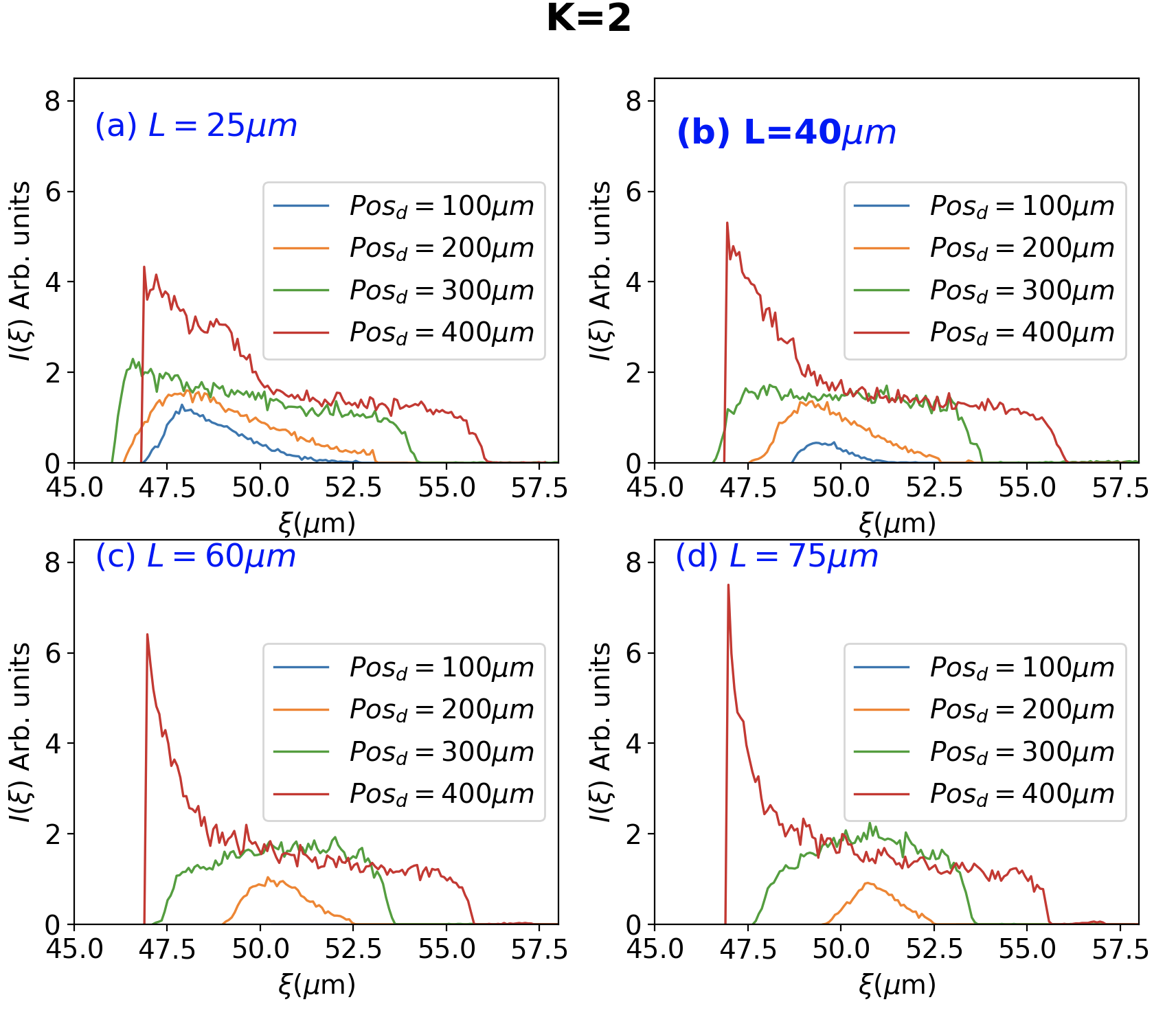}
\caption{\label{fig: I(xi) combo} Beam sliced currents when the ratio $K=2$ with different downramp lengths $L$ and downramp positions $\text{Pos}_d$.}
\end{figure}
To investigate the effect of the density gradient on the beam parameters, we performed additional simulations for different dowmramp length $L$. Figure \ref{fig: I(xi) combo} reports on the beam current distributions for four different density gradients values and four different downramp positions $\text{Pos}_d$. The results indicate that the more steepen the downramp (smaller $L$) is, the easier the injection becomes. Note that, no injection occurs  for $L=60\mu m$ and $L=75\mu m$ when the $Pos_{d}$ is set as 100 $\mu$m. The steepness of the downramp and the linearity in the space of $\xi$ and $z_i$ gives possibilities to obtaining the desired beam current $I(\xi)$ by adjusting downramp length and placing downramp region according to the laser intensity evolution.

\begin{figure}
\includegraphics[width=0.6\textwidth]{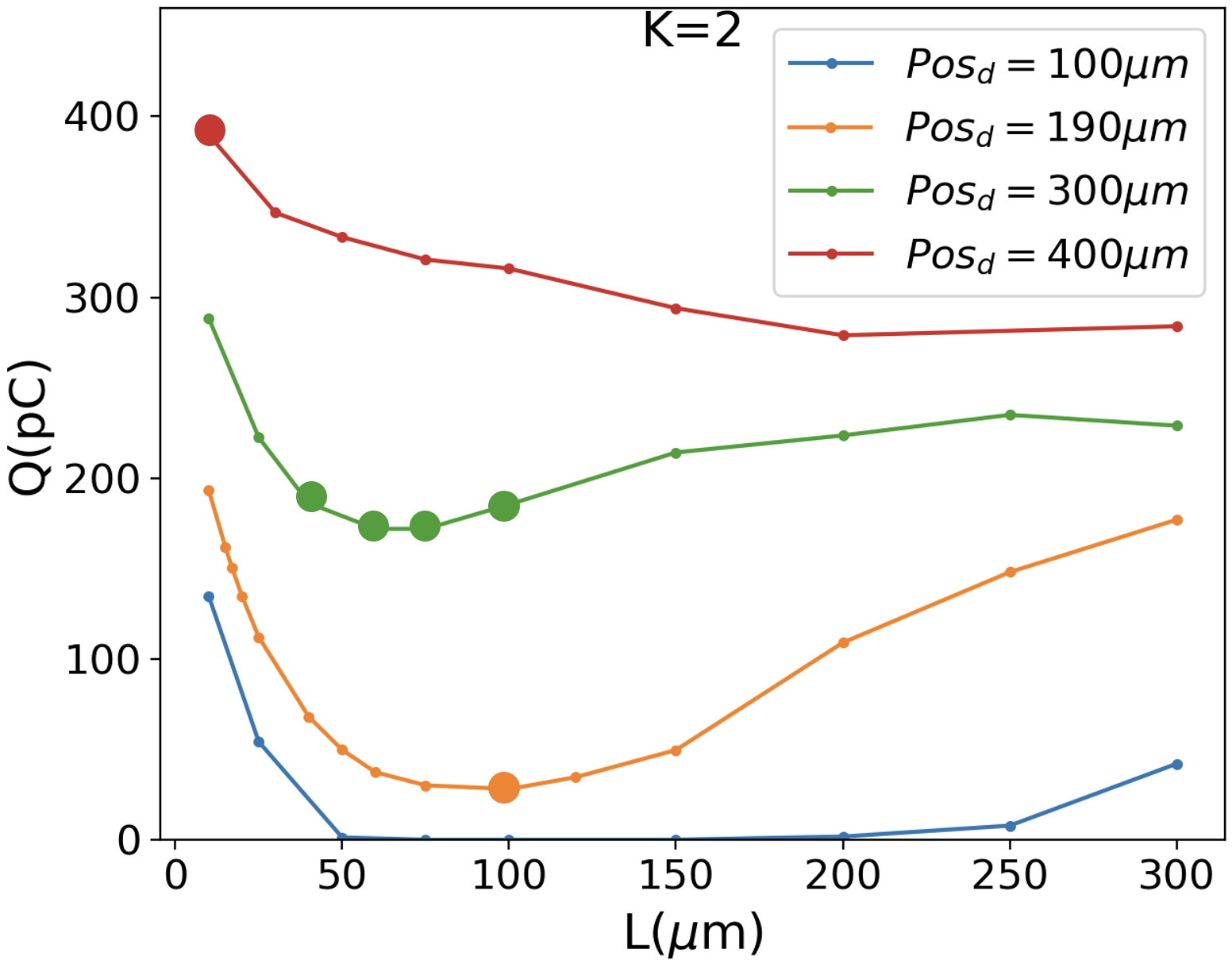}
\caption{\label{fig: Q scans} Beam charges $Q$ when the ratio $K=2$ with different downramp lengths $L$ and downramp positions $\text{Pos}_d$.}
\end{figure}

It is interesting to calculate total injected charge $Q=1/c \int_{\text{beam}} I(\xi)d\xi$ that each set of plasma parameters produces. In Figure \ref{fig: Q scans}, the beam charge $Q$ varies according to the downramp length $L$ for four downramp positions $\text{Pos}_d$ where three of them are previously considered. The large dots data indicated the cases where energy performances are optimal. These cases will be discussed in section \ref{sec:I(z) and energy spread}.

Simulations (not reported here) were also performed with different $K$. For situations where a complete blow-out is not achieved at the beginning of the downramp, we obtain similar $I(\xi)$ for same downramp steepness $K/L$ when controlling $a_0$ across the downramp region in adjusting $\text{Pos}_d$. For the full blow-out situations across the downramp region, the injected beam is observed to have a longer duration with higher $K$, which potentially increase the beam charge $Q$. The reason is that the bubble size in the high-density region is smaller for higher $K$ ($n_{low}$ is fixed) which makes the head of injected electron bunch is closer to the laser driver while its tail stays at the rear of the bubble. But $Q$ is also limited by the acceleration capacity of the structure after downramp as discussed in section \ref{sec:limit}.

\section{\label{sec:I(z) and energy spread} Control of beam energy evolution and optimization of beam parameters}

As is stated in section \ref{sec: tailoring}, the shape of the beam current $I(\xi)$ has decisive influence on beam energy evolution through propagation. 

\begin{figure}[ht!]
\includegraphics[width=0.7\textwidth]{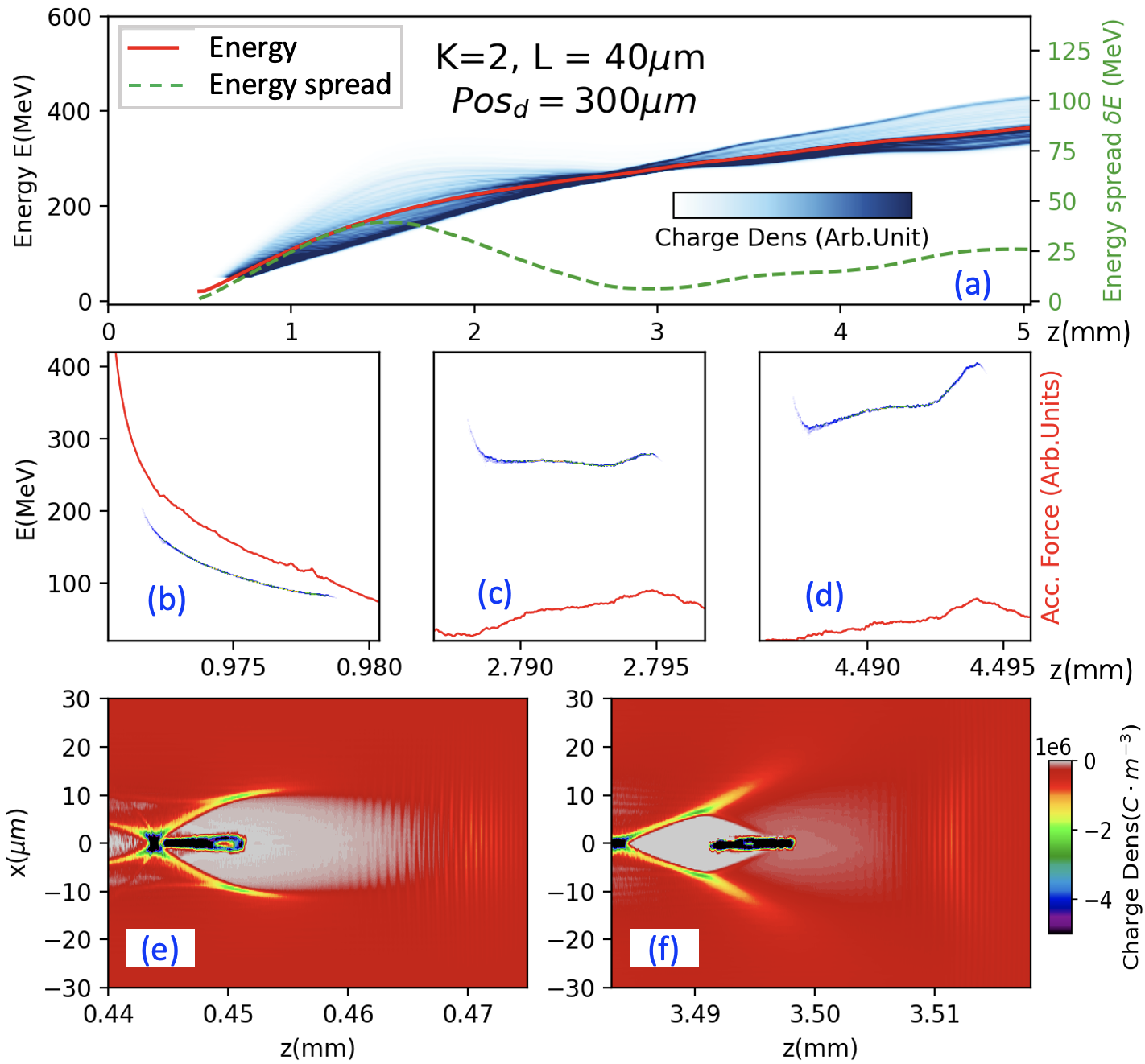}
\caption{\label{fig: L40 dE evol exmp} The evolution of beam energy over propagation for simulation where $K=2$, $\text{Pos}_d=300\mu m$ and $L=40\mu m$ is plotted in the sub-figure (d). The beam profile in longitudinal phase space for different propagation distance is plotted in sub-figures (a), (b) and (c), where the red lines are the on-axis acceleration forces. Wakefield of two different propagation distances are illustrated in sub-figure (e) and (f).}
\end{figure}

A typical simulation with initial plasma density parameters $K=2$, $\text{Pos}_d=300\mu m$ and $L=40\mu m$ is presented here for the production of mono-energetic beams with energy of hundreds of MeVs. The electron beam energy evolution is shown in figure \ref{fig: L40 dE evol exmp}(d),where energy spectrum for the injected bunch over propagation is shown as color-map, the beam mean energy is plotted as red solid line of which the scale refers to ticks on the left vertical axis and beam absolute energy spread $\delta E$ is plotted as green dashed line of which the scale refers to ticks on the right vertical axis. The $I(\xi)$ of the injected beam in this simulation corresponds to the green line in figure \ref{fig: I(xi) 40}. 

Through propagation, beam energy spread reaches its minimum before re-increasing over propagation. For understanding the reason of such energy evolution, the electron beam density distributions in longitudinal phase space ($z-E$ plane) at different propagation distance, as well as the acceleration force that electrons are submitted, are plotted at in sub-figure (a), (b) and (c). It is observed, through the very thin beam profiles in longitudinal phase spaces, that large energy spreads are due to the energy chirp whereas the sliced energy spreads stays very low throughout the whole propagation. Beam chirps negatively (tail has higher energy than the head of the bunch) at the beginning of the simulation, then flattens over propagation before showing nearly-horizontal shape (the head and tail of the beam holds nearly same energy). The nearly-horizontal shape is where the minimum energy spread is found. Then the chirp turns positive and continuously increases till the end of the simulation and energy spread increases correspondingly. The optimised propagation distance for obtaining good beam properties is thus around zones where the minimum energy spread is obtained (for longer propagation distance the beam energy does not increase sensitively). Here, we focus on three elements that characterise the beam energy evolution : 1) Minimum energy spread $\min (\delta E)$; 2) The beam mean energy where minimum energy spread reaches $E(\min(\delta E)$; 3) The rate at which $\delta E$ increases after its minima $\text{d}(\delta E)/\text{d}z$.

The evolution of beam profile in longitudinal phase space is a result of the evolution of acceleration force $F_\text{acc}(z)$ received by the beam at different longitudinal position . It is a combined effect of the laser-driven wakefield and the beam self-loaded wakefield. Shortly after the bunch is injected, the acceleration force that the tail of the beam receives is stronger than its front, which results into the negative chirp. When the wakefield structure propagates in plasma, the acceleration force weakens with the laser depletion. Accordingly, the beam self-loaded wakefield which gives a deceleration force dominates at the rear of the beam. The wakefield structure shortly after downramp and near the propagation distance of 3.5 mm are illustrated respectively in sub-figure \ref{fig: L40 dE evol exmp}.e) and \ref{fig: L40 dE evol exmp}.f) respectively. It shows that the wakefiled contributed by the laser (structure before the beam head, to the right of the beam position in the figure) becomes much weaker after few mm propagation and thus the beam self-loaded wakefield starts to be dominant. 

As explained in \cite{Tzoufras2008}, one can shape the acceleration force by modifying beam current $I(\xi)$. Since the beam current profile $I(\xi)$ relies on the three parameters of the density transition region as mentioned in section \ref{sec: tailoring}, the beam energy performance can thus be optimized by tuning the three parameters of plasma profiles: $K$, $L$, and $\text{Pos}_d$.

\begin{figure*}[ht!]
\includegraphics[width=0.85\textwidth]{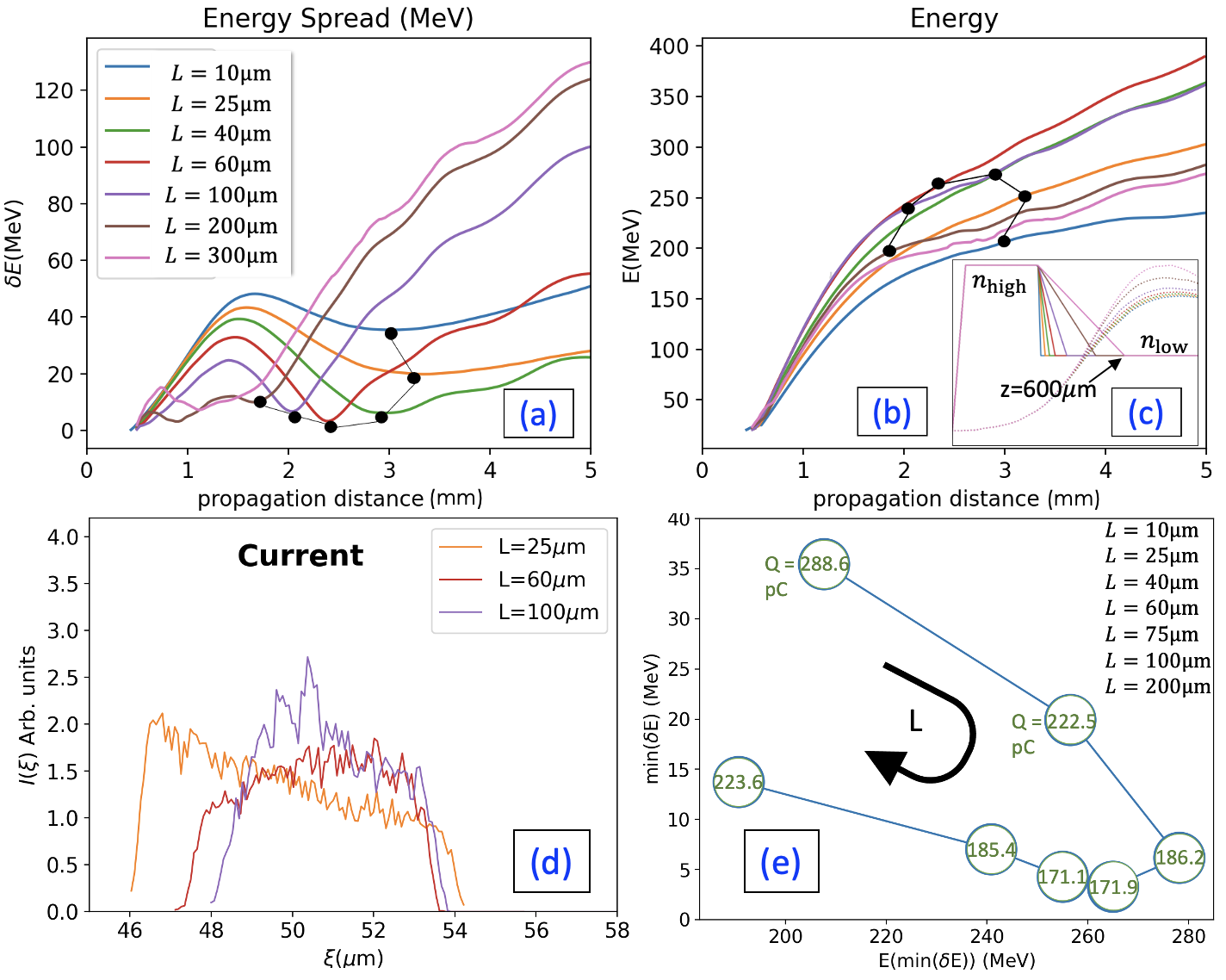}
\caption{\label{fig:E sans L300} The evolution of $\delta E$ and $E$ corresponding to different $L$ are plotted in sub-figure (a) and (b) when $\text{Pos}_d=300\mu m$ and $K=2$. Black dots mark the optimized zones. The plasma density and laser $a_0$ evolution are illustrated in the inset figure (c). Three beam currents are illustrated in sub-figure (d). Relationship of the minimum energy spread, energy, and injected beam charge for different values of downramp gradient $L$, is shown in sub figure (e), with for each points the encircled charge values.}
\end{figure*}

A series of simulations are performed for mapping the relationship of plasma parameters and the evolution of beam energy. The first group of simulations refers to the green line in figure \ref{fig: Q scans}, where the ratio $K=2$ and downramp position $\text{Pos}_d=300\mu m$. Only the downramp length $L$ varies. The evolution of energy spread $\delta E$ over propagation for several cases in this group are plotted in Figure \ref{fig:E sans L300}.a). The profiles of energy evolution for the whole beam are plotted in Figure \ref{fig:E sans L300}.b). Black dots are outlines of the zones of interests where minimum energy spreads $\min\delta E$ are reached. Plasma density profiles and $a_0$ evolution are illustrated in the inset-figure (c). 

The minimum energy spread is very sensitive to the downramp gradient $L$, with values that are larger for smaller $L$. The value $\min(\delta E)$ reduces with larger $L$ and reaches its lowest value for the performed simulation in this group when $L=60\mu m$. When $L$ continues increasing, the beam energy obtained at the optimized zone for beam delivery $E(\min(\delta E))$ decreases, and the speed at which $\delta E$ evolves $\text{d}(\delta E)/\text(d)z$ after the zone of delivery is higher which makes the system more sensitive to be experimentally controlled. 


We are furthermore interested in the relationship between achievable $\delta E$, $E$ and $Q$, which we consider as three main parameters that characterize the injected beam and reported in Figure. 7(e). The downwramp length $L$ increases with the flesh direction. The beam charge $Q$ in pico-Coulomb is written inside each dot, of which the horizontal position is $E(\min(\delta E))$ and the vertical one $\min(\delta E)$. It can be seen that the achievable minimum energy spread coincides with the attainable maximum energies (right-most dots in sub-figure (e) are also the lowest), which is favorable for the production of high-energy mono-energetic beams. Table \ref{table: Only one table} presents the final beam parameters of  four downramp length $L$ where $K=2$ and $\text{Pos}_d = 300 \mu m$ is fixed, which are considered to be optimum in the frame of obtaining high-energy mono-energetic beams, i.e. the smallest $\delta E/E$ values.

\begin{figure*}[ht!]
\includegraphics[width=0.7\textwidth]{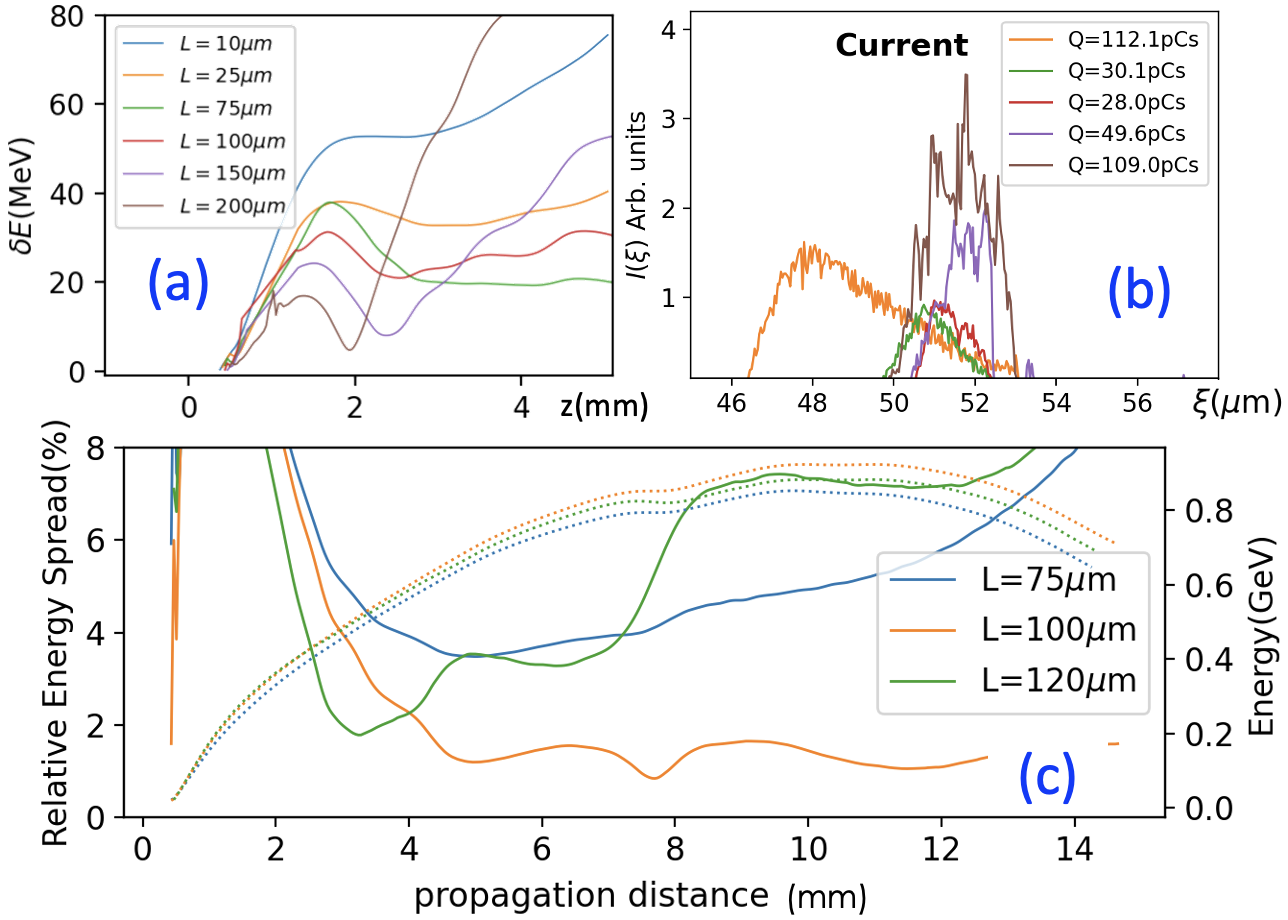}
\caption{\label{fig:E sans L200} Energy spread (a) as a function of the propagation distance for six values of the downramp gradients; beam current (b) for corresponding cases; relative energy spread and energy as a function of the propagation distance for a set of four downramp gradient values (c).}
\end{figure*}

\begin{table*}[t]
\begin{center}{}
\resizebox{0.7\textwidth}{!}{
\begin{tabular}{l|p{1.5cm}p{1.5cm}p{1.5cm}|p{1.5cm}p{1.5cm}p{1.5cm}p{1.5cm}}
 & $\text{Pos}_d$($\mu$m) & $L$($\mu$m) & $K$ & Q(pC) & $\delta E$(MeV) & $E$(MeV) & $\delta E/E$(\%) \\ 
 \hline
\multirow{4}{5em}{Sim.gp. 1}  & 300 & 40 & 2 & 186.2 & 6.20 & 278.14 & 2.2 \\ 
  & 300 & 60 & 2 & 171.5 & 3.31 & 265.02 & 1.2\\ 
  & 300 & 75 & 2 & 171.1 & 4.32 & 255.03 & 1.7\\ 
  & 300 & 100 & 2 & 185.4 & 7.05 & 240.82 & 2.9\\
 \hline
 Sim.gp. 2 & 190 & 100 & 2 & 28.0 & 10.22 & 993.2 & 1.03 \\
 \hline
 Sim.gp. 3 & 400 & 10 & 2 & 390.1 & 7.70 & 144.5 & 5.3 \\
 \hline
\end{tabular}
}
\caption{Plasma parameters and final beam parameters for the optimum cases that give the smallest $\delta E/E$ for each group of simulations.}
\label{table: Only one table}
\end{center}
\end{table*}

The second group of simulations refers to orange line in figure \ref{fig: Q scans}, where $K=2$ and $\text{Pos}_d=190 \mu$m. Beam parameters evolution for some cases in this group are plotted in figure \ref{fig:E sans L200}. It is observed that $\delta E$ keeps stable till 5 mm propagation length for the case where $L=100\mu m$. Further investigations with longer propagation distance are performed for the simulations of interest where $\delta E$ be conserved. The relative energy spreads, which characterize better beam properties than the absolute ones are plotted in sub-figure (c). The relative energy spreads are preserved and reach minimum where we consider the optimized zones of beam delivery for this group. Beam energies reach approximately their maximum around the zones of beam deliveries and rise up to approximately 1GeV while low relative energy spread (inferior to 2.5\%) is obtained. Beam currents for four cases are plotted in sub-figure (b), of which the color corresponds to sub-figure (a). The current shape for this simulation group are pulse-like. Similar to the fist group of simulations, the currents move more to the rear of the bubble with larger $L$ and the beam length reduces significantly. The cases where energy spreads are conserved are only for cases with weak beam current, thus, low charge $Q$. The possible preservation of energy spread is due to the short beam lengths and weak current strength, because of which the beam loading is not significant and the wakefield driven by the laser is dominant throughout the propagation. Laser is self-guided \cite{LU2007LaserSelfGuiding} and provides a stable wakefield for long propagation distance. The beam charge $Q$ for the cases of interest are around 30pC. The optimized case of this group is presented in table \ref{table: Only one table}.

\begin{figure}[ht!]
\includegraphics[width=0.85\textwidth]{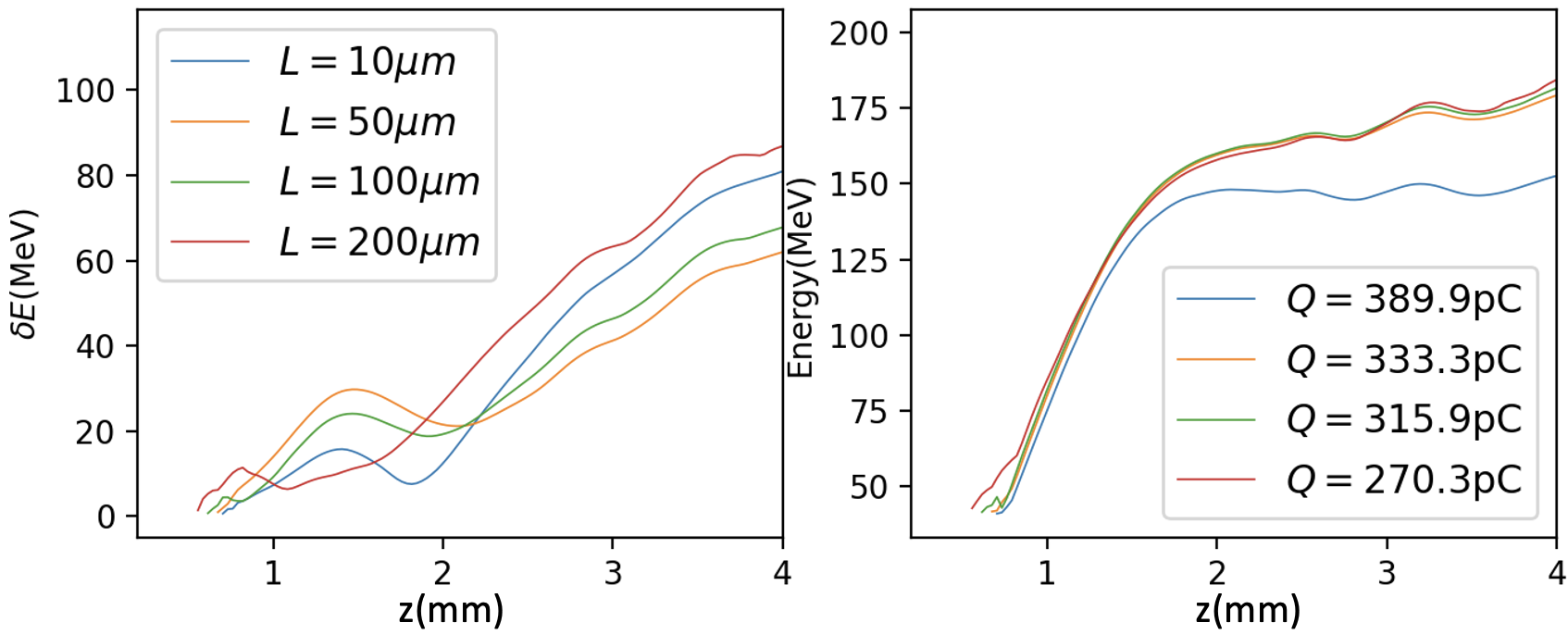}
\caption{\label{fig:E sans L400} Energy spread and energy evolution patterns at $\text{Pos}_d=400\mu m$ and for four downramp density gradient values.}
\end{figure}

We then focus on the third group of simulations, which refers to the red line in figure \ref{fig: Q scans}. The energy evolution for several cases in this group are presented in figure \ref{fig:E sans L400}. Beams are barely accelerated after reaching 150 MeV. Similar to the former groups of simulations, the optimized beam delivery zone is where the minimum energy spread reaches. Among all performed simulations in this group, the most mono-energetic case is found with the smallest $L$, where the injected beam charge approaches 390pC.

\section{\label{sec:limit}Limits for acceleration capacities}
\begin{figure*}[ht!]
\includegraphics[width=0.8\textwidth]{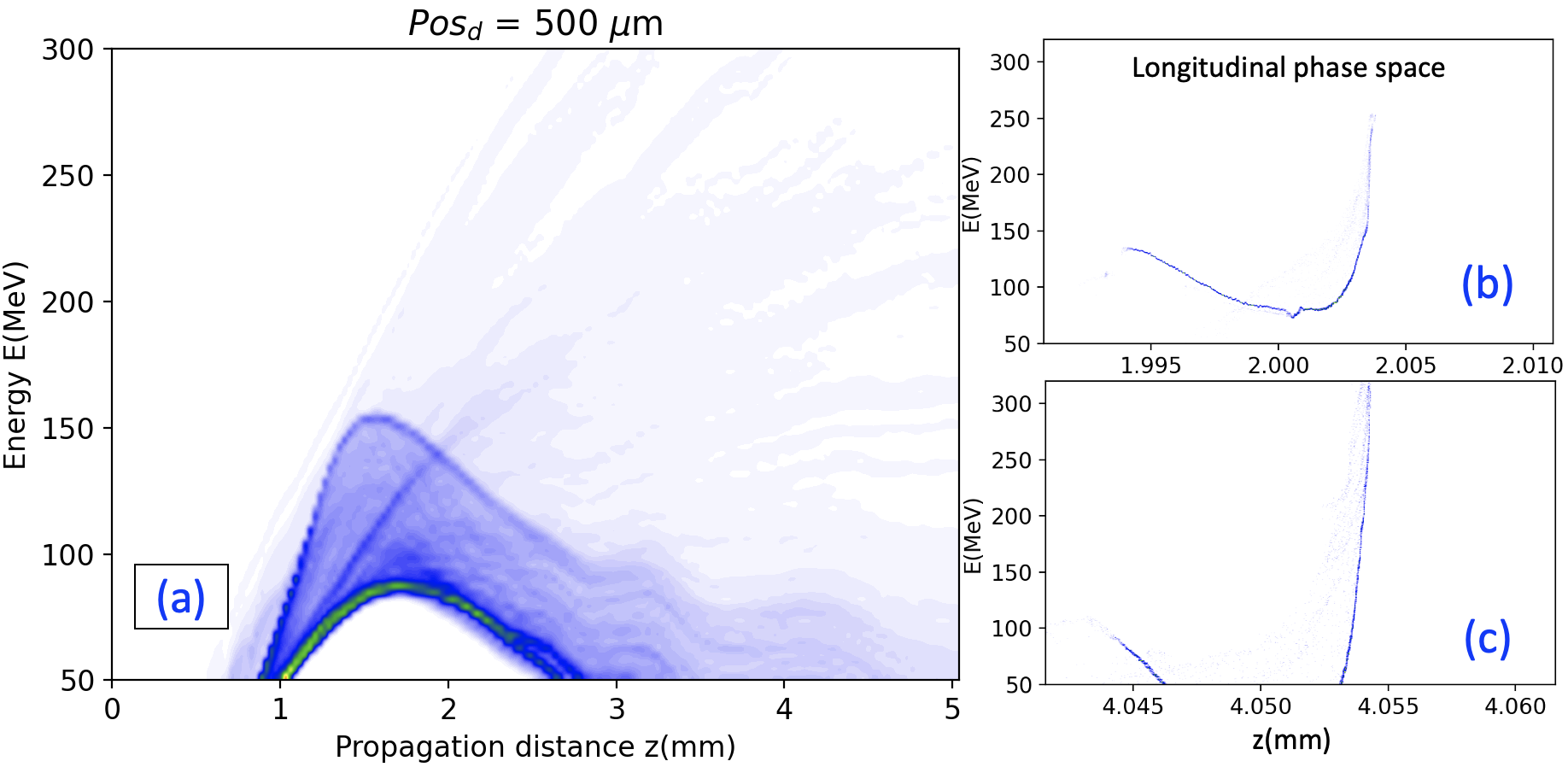}
\caption{\label{fig:Decc lim} Energy spectrum and illustration of longitudinal phase space for two different propagation distances, and with initial plasma parameters $K=2$, $L=25\mu m$ and $\text{Pos}_d = 500\mu m$. Significant particle loss is observed after 3 mm of propagation.}
\end{figure*}

One can expect increasing $a_0$ across downramp region for obtaining higher beam charge. However, the beam charge is limited by the wakefield amplitude, depends on the laser evolution after the downramp and the plasma density $n_{\text{low}}$. We call here this limit the \textit{deceleration limit}. 

An energy spectrum for a simulation which exceeds the deceleration limit is shown in figure \ref{fig:Decc lim}. When the beam current is high enough and accordingly the total beam charge, the decelerating wakefield loaded by the beam itself becomes too strong that the majority of particles in the beam is decelerated. Figure \ref{fig:Decc lim} shows the energy spectrum and the beam profile in longitudinal phase space for the described situation. Initial plasma parameters of this simulation are $K=2$, $\text{Pos}_d=435\mu m$ and $L=25\mu m$. The injected charge is 394.4pC. 

\begin{figure}[ht!]
\includegraphics[width=0.48\textwidth]{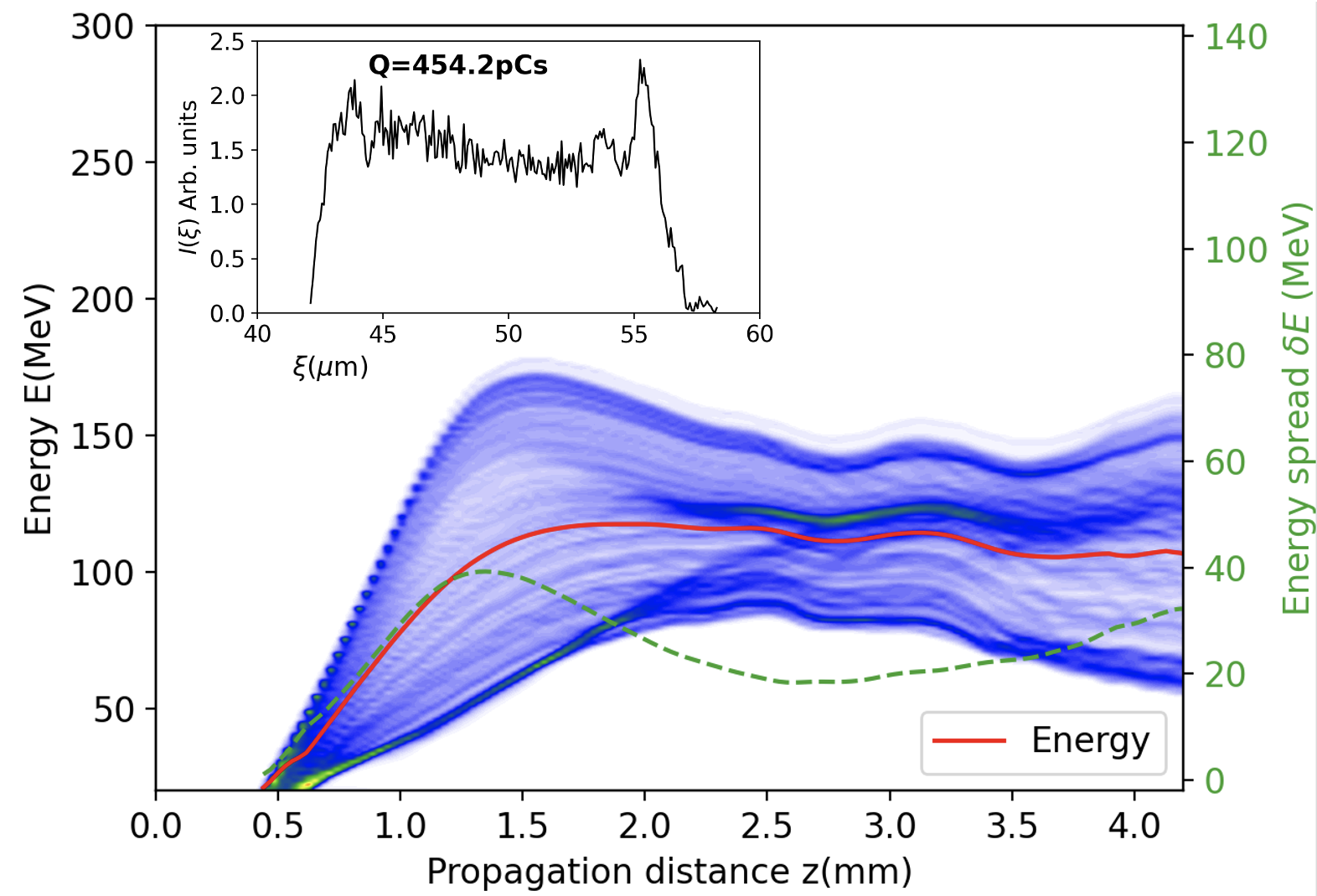}
\caption{\label{fig:E_maxQ} The energy spectrum for the simulation with initial palsma parameters $K=4$, $L=25\mu m$ and $\text{Pos}_d = 225\mu m$. The beam current for this simulation is plotted in the inset figure.}
\end{figure}

The maximum injected charge obtained among all performed simulations for the constant $n_{\text{low}} =2.5\times 10^{18}cm^{-3}$ plasma density located after downramp is 454.2pC. For this case, $K=4$, $\text{Pos}_d=225\mu m$ and $L=25\mu m$. The beam current is nearly flat which makes the energy spread relatively stable over propagation as shown in figure \ref{fig:E_maxQ}. 

Increasing more $K$ does not help to reach higher beam charge because the bunch length cannot exceed the acceleration and focusing part of bubble structure. One can expect lowering $n_{\text{low}}$ for a larger bubble structure for being able to accelerate longer bunches. 

\section{Conclusion}
In this article, a detail and comprehensive study on obtaining high quality electron beams using density downramp injection technique in a one-stage Laser Plasma Wakefield Accelerator is reported. It is shown how the plasma target parameters are affecting the beam performances.

It is understood that the beam current $I(\xi)$ is of decisive importance not only for obtaining desirable beam charge but also in the beam energy evolution. This current profile can be tailored by positioning the plasma downramp region in regard to the local laser vector potential $a_0$. Different $I(\xi)$ give result to different energy evolution patterns. The achievable beam charge for a fixed $n_{\text{low}} =2.5\times 10^{18}cm^{-3}$ varies from 0 to 450 pC. The beams with high-most charge and variable current shapes can be potentially tuned and adapted for offering driver beams for high-transformer ratio PWFA \cite{Gotzfried2020,Corde2015}. 

For injected charges at the scale of hundreds of pico-Coulomb, a minimum energy spread $\min(\delta E)$ is found as a result of the rivalry between self-loaded decelerating field and the evolution accelerating wakefield driven by the laser over propagation. Optimum zones of beam deliveries are where $\min(\delta E)$ is reached. Beams with charge over 170pC, energy in the 200MeV range and less than 2\% relative energy spread are obtained, and that fit nicely the requirement for VHEE application \cite{Glinec2006,Fuchs_2009,Malka2008}.

For cases where relatively low charge is injected, conservation of energy spread over long propagation distance is observed which permit the beam to reach very high energy (approximately 1GeV) with less than 2\% energy spread. Such cases can be further tuned and adapted to FEL applications that has been recently demonstrated on the basis of \cite{Wang2021FEL}.

In this paper we give general ideas about how to tune parameters and their ranges. Further optimisations are required orienting to different applications and available experimental facilities.


\begin{acknowledgments}
This work was supported by the \textit{Fondation Jacques Toledano}, the \textit{Schwartz/Reisman Center for Intense Laser Physics}, and by the ERC \textit{PoC Vherapy} and the EIC \textit{ebeam4therapy} grants.
\end{acknowledgments}




\nocite{*}

\bibliography{apssamp}

\begin{thebibliography}{30}%
\makeatletter
\providecommand \@ifxundefined [1]{%
 \@ifx{#1\undefined}
}%
\providecommand \@ifnum [1]{%
 \ifnum #1\expandafter \@firstoftwo
 \else \expandafter \@secondoftwo
 \fi
}%
\providecommand \@ifx [1]{%
 \ifx #1\expandafter \@firstoftwo
 \else \expandafter \@secondoftwo
 \fi
}%
\providecommand \natexlab [1]{#1}%
\providecommand \enquote  [1]{``#1''}%
\providecommand \bibnamefont  [1]{#1}%
\providecommand \bibfnamefont [1]{#1}%
\providecommand \citenamefont [1]{#1}%
\providecommand \href@noop [0]{\@secondoftwo}%
\providecommand \href [0]{\begingroup \@sanitize@url \@href}%
\providecommand \@href[1]{\@@startlink{#1}\@@href}%
\providecommand \@@href[1]{\endgroup#1\@@endlink}%
\providecommand \@sanitize@url [0]{\catcode `\\12\catcode `\$12\catcode
  `\&12\catcode `\#12\catcode `\^12\catcode `\_12\catcode `\%12\relax}%
\providecommand \@@startlink[1]{}%
\providecommand \@@endlink[0]{}%
\providecommand \url  [0]{\begingroup\@sanitize@url \@url }%
\providecommand \@url [1]{\endgroup\@href {#1}{\urlprefix }}%
\providecommand \urlprefix  [0]{URL }%
\providecommand \Eprint [0]{\href }%
\providecommand \doibase [0]{https://doi.org/}%
\providecommand \selectlanguage [0]{\@gobble}%
\providecommand \bibinfo  [0]{\@secondoftwo}%
\providecommand \bibfield  [0]{\@secondoftwo}%
\providecommand \translation [1]{[#1]}%
\providecommand \BibitemOpen [0]{}%
\providecommand \bibitemStop [0]{}%
\providecommand \bibitemNoStop [0]{.\EOS\space}%
\providecommand \EOS [0]{\spacefactor3000\relax}%
\providecommand \BibitemShut  [1]{\csname bibitem#1\endcsname}%
\let\auto@bib@innerbib\@empty
\bibitem [{\citenamefont {Faure}\ \emph {et~al.}(2004)\citenamefont {Faure},
  \citenamefont {Glinec}, \citenamefont {Pukhov}, \citenamefont {Kiselev},
  \citenamefont {Gordienko}, \citenamefont {Lefebvre}, \citenamefont
  {Rousseau}, \citenamefont {Burgy},\ and\ \citenamefont {Malka}}]{Faure2004}%
  \BibitemOpen
  \bibfield  {author} {\bibinfo {author} {\bibfnamefont {J.}~\bibnamefont
  {Faure}}, \bibinfo {author} {\bibfnamefont {Y.}~\bibnamefont {Glinec}},
  \bibinfo {author} {\bibfnamefont {A.}~\bibnamefont {Pukhov}}, \bibinfo
  {author} {\bibfnamefont {S.}~\bibnamefont {Kiselev}}, \bibinfo {author}
  {\bibfnamefont {S.}~\bibnamefont {Gordienko}}, \bibinfo {author}
  {\bibfnamefont {E.}~\bibnamefont {Lefebvre}}, \bibinfo {author}
  {\bibfnamefont {J.~P.}\ \bibnamefont {Rousseau}}, \bibinfo {author}
  {\bibfnamefont {F.}~\bibnamefont {Burgy}},\ and\ \bibinfo {author}
  {\bibfnamefont {V.}~\bibnamefont {Malka}},\ }\bibfield  {title} {\enquote
  {\bibinfo {title} {A laser-plasma accelerator producing monoenergetic
  electron beams},}\ }\href {https://www.nature.com/articles/nature02963}
  {\bibfield  {journal} {\bibinfo  {journal} {Nature (London)}\ }\textbf
  {\bibinfo {volume} {431}},\ \bibinfo {pages} {541--544} (\bibinfo {year}
  {2004})}\BibitemShut {NoStop}%
\bibitem [{\citenamefont {Geddes}\ \emph {et~al.}(2004)\citenamefont {Geddes},
  \citenamefont {Toth}, \citenamefont {van Tilborg}, \citenamefont {Esarey},
  \citenamefont {Schroeder}, \citenamefont {Bruhwiler}, \citenamefont {Nieter},
  \citenamefont {Cary},\ and\ \citenamefont {Leemans}}]{Geddes2004}%
  \BibitemOpen
  \bibfield  {author} {\bibinfo {author} {\bibfnamefont {C.~G.~R.}\
  \bibnamefont {Geddes}}, \bibinfo {author} {\bibfnamefont {C.}~\bibnamefont
  {Toth}}, \bibinfo {author} {\bibfnamefont {J.}~\bibnamefont {van Tilborg}},
  \bibinfo {author} {\bibfnamefont {E.}~\bibnamefont {Esarey}}, \bibinfo
  {author} {\bibfnamefont {C.~B.}\ \bibnamefont {Schroeder}}, \bibinfo {author}
  {\bibfnamefont {D.}~\bibnamefont {Bruhwiler}}, \bibinfo {author}
  {\bibfnamefont {C.}~\bibnamefont {Nieter}}, \bibinfo {author} {\bibfnamefont
  {J.}~\bibnamefont {Cary}},\ and\ \bibinfo {author} {\bibfnamefont {W.~P.}\
  \bibnamefont {Leemans}},\ }\bibfield  {title} {\enquote {\bibinfo {title}
  {High-quality electron beams from a laser wakefield accelerator using
  plasma-channel guiding},}\ }\href
  {https://www.nature.com/articles/nature02900} {\bibfield  {journal} {\bibinfo
   {journal} {Nature (London)}\ }\textbf {\bibinfo {volume} {431}},\ \bibinfo
  {pages} {538--541} (\bibinfo {year} {2004})}\BibitemShut {NoStop}%
\bibitem [{\citenamefont {Mangles}\ \emph {et~al.}(2004)\citenamefont
  {Mangles}, \citenamefont {Murphy}, \citenamefont {Najmudin}, \citenamefont
  {Thomas}, \citenamefont {Collier}, \citenamefont {Dangor}, \citenamefont
  {Divall}, \citenamefont {Foster}, \citenamefont {Gallacher}, \citenamefont
  {Hooker}, \citenamefont {Jaroszynski}, \citenamefont {Langley}, \citenamefont
  {Mori}, \citenamefont {Norreys}, \citenamefont {Tsung}, \citenamefont
  {Viskup}, \citenamefont {Walton},\ and\ \citenamefont
  {Krushelnick}}]{Mangles2004}%
  \BibitemOpen
  \bibfield  {author} {\bibinfo {author} {\bibfnamefont {S.~P.~D.}\
  \bibnamefont {Mangles}}, \bibinfo {author} {\bibfnamefont {C.~D.}\
  \bibnamefont {Murphy}}, \bibinfo {author} {\bibfnamefont {Z.}~\bibnamefont
  {Najmudin}}, \bibinfo {author} {\bibfnamefont {A.~G.~R.}\ \bibnamefont
  {Thomas}}, \bibinfo {author} {\bibfnamefont {J.~L.}\ \bibnamefont {Collier}},
  \bibinfo {author} {\bibfnamefont {A.~E.}\ \bibnamefont {Dangor}}, \bibinfo
  {author} {\bibfnamefont {E.~J.}\ \bibnamefont {Divall}}, \bibinfo {author}
  {\bibfnamefont {P.~S.}\ \bibnamefont {Foster}}, \bibinfo {author}
  {\bibfnamefont {J.~G.}\ \bibnamefont {Gallacher}}, \bibinfo {author}
  {\bibfnamefont {C.~J.}\ \bibnamefont {Hooker}}, \bibinfo {author}
  {\bibfnamefont {D.~A.}\ \bibnamefont {Jaroszynski}}, \bibinfo {author}
  {\bibfnamefont {A.~J.}\ \bibnamefont {Langley}}, \bibinfo {author}
  {\bibfnamefont {W.~B.}\ \bibnamefont {Mori}}, \bibinfo {author}
  {\bibfnamefont {P.~A.}\ \bibnamefont {Norreys}}, \bibinfo {author}
  {\bibfnamefont {F.~S.}\ \bibnamefont {Tsung}}, \bibinfo {author}
  {\bibfnamefont {R.}~\bibnamefont {Viskup}}, \bibinfo {author} {\bibfnamefont
  {B.~R.}\ \bibnamefont {Walton}},\ and\ \bibinfo {author} {\bibfnamefont
  {K.}~\bibnamefont {Krushelnick}},\ }\bibfield  {title} {\enquote {\bibinfo
  {title} {Monoenergetic beams of relativistic electrons from intense
  laser-plasma interactions},}\ }\href
  {https://www.nature.com/articles/nature02939} {\bibfield  {journal} {\bibinfo
   {journal} {Nature (London)}\ }\textbf {\bibinfo {volume} {431}},\ \bibinfo
  {pages} {535--538} (\bibinfo {year} {2004})}\BibitemShut {NoStop}%
\bibitem [{\citenamefont {Tajima}\ and\ \citenamefont
  {Dawson}(1979)}]{Tajima1979}%
  \BibitemOpen
  \bibfield  {author} {\bibinfo {author} {\bibfnamefont {T.}~\bibnamefont
  {Tajima}}\ and\ \bibinfo {author} {\bibfnamefont {J.~M.}\ \bibnamefont
  {Dawson}},\ }\bibfield  {title} {\enquote {\bibinfo {title} {Laser electron
  accelerator},}\ }\href {https://doi.org/10.1103/PhysRevLett.43.267}
  {\bibfield  {journal} {\bibinfo  {journal} {Phys. Rev. Lett.}\ }\textbf
  {\bibinfo {volume} {43}},\ \bibinfo {pages} {267--270} (\bibinfo {year}
  {1979})}\BibitemShut {NoStop}%
\bibitem [{\citenamefont {Malka}(2012)}]{Malka2012}%
  \BibitemOpen
  \bibfield  {author} {\bibinfo {author} {\bibfnamefont {V.}~\bibnamefont
  {Malka}},\ }\bibfield  {title} {\enquote {\bibinfo {title} {Laser plasma
  accelerators},}\ }\href {https://doi.org/10.1063/1.3695389} {\bibfield
  {journal} {\bibinfo  {journal} {Physics of Plasmas}\ }\textbf {\bibinfo
  {volume} {19}},\ \bibinfo {pages} {055501} (\bibinfo {year} {2012})},\
  \Eprint {https://arxiv.org/abs/https://doi.org/10.1063/1.3695389}
  {https://doi.org/10.1063/1.3695389} \BibitemShut {NoStop}%
\bibitem [{\citenamefont {Bulanov}\ \emph {et~al.}(1998)\citenamefont
  {Bulanov}, \citenamefont {Naumova}, \citenamefont {Pegoraro},\ and\
  \citenamefont {Sakai}}]{Bulanov1998}%
  \BibitemOpen
  \bibfield  {author} {\bibinfo {author} {\bibfnamefont {S.}~\bibnamefont
  {Bulanov}}, \bibinfo {author} {\bibfnamefont {N.}~\bibnamefont {Naumova}},
  \bibinfo {author} {\bibfnamefont {F.}~\bibnamefont {Pegoraro}},\ and\
  \bibinfo {author} {\bibfnamefont {J.}~\bibnamefont {Sakai}},\ }\bibfield
  {title} {\enquote {\bibinfo {title} {Particle injection into the wave
  acceleration phase due to nonlinear wake wave breaking},}\ }\href
  {https://doi.org/10.1103/PhysRevE.58.R5257} {\bibfield  {journal} {\bibinfo
  {journal} {Phys. Rev. E}\ }\textbf {\bibinfo {volume} {58}},\ \bibinfo
  {pages} {R5257--R5260} (\bibinfo {year} {1998})}\BibitemShut {NoStop}%
\bibitem [{\citenamefont {Thaury}\ \emph {et~al.}(2015)\citenamefont {Thaury},
  \citenamefont {Guillaume}, \citenamefont {Lifschitz}, \citenamefont
  {Ta~Phuoc}, \citenamefont {Hansson}, \citenamefont {Grittani}, \citenamefont
  {Gautier}, \citenamefont {Goddet}, \citenamefont {Tafzi}, \citenamefont
  {Lundh},\ and\ \citenamefont {Malka}}]{Thaury2015}%
  \BibitemOpen
  \bibfield  {author} {\bibinfo {author} {\bibfnamefont {C.}~\bibnamefont
  {Thaury}}, \bibinfo {author} {\bibfnamefont {E.}~\bibnamefont {Guillaume}},
  \bibinfo {author} {\bibfnamefont {A.}~\bibnamefont {Lifschitz}}, \bibinfo
  {author} {\bibfnamefont {K.}~\bibnamefont {Ta~Phuoc}}, \bibinfo {author}
  {\bibfnamefont {M.}~\bibnamefont {Hansson}}, \bibinfo {author} {\bibfnamefont
  {G.}~\bibnamefont {Grittani}}, \bibinfo {author} {\bibfnamefont
  {J.}~\bibnamefont {Gautier}}, \bibinfo {author} {\bibfnamefont {J.~P.}\
  \bibnamefont {Goddet}}, \bibinfo {author} {\bibfnamefont {A.}~\bibnamefont
  {Tafzi}}, \bibinfo {author} {\bibfnamefont {O.}~\bibnamefont {Lundh}},\ and\
  \bibinfo {author} {\bibfnamefont {V.}~\bibnamefont {Malka}},\ }\bibfield
  {title} {\enquote {\bibinfo {title} {Shock assisted ionization injection in
  laser-plasma accelerators},}\ }\href {https://doi.org/10.1038/srep16310}
  {\bibfield  {journal} {\bibinfo  {journal} {Scientific Reports}\ }\textbf
  {\bibinfo {volume} {5}},\ \bibinfo {pages} {16310} (\bibinfo {year}
  {2015})}\BibitemShut {NoStop}%
\bibitem [{\citenamefont {Buck}\ \emph
  {et~al.}(2013{\natexlab{a}})\citenamefont {Buck}, \citenamefont {Wenz},
  \citenamefont {Xu}, \citenamefont {Khrennikov}, \citenamefont {Schmid},
  \citenamefont {Heigoldt}, \citenamefont {Mikhailova}, \citenamefont
  {Geissler}, \citenamefont {Shen}, \citenamefont {Krausz}, \citenamefont
  {Karsch},\ and\ \citenamefont {Veisz}}]{Buck2013ShockInjectionExp}%
  \BibitemOpen
  \bibfield  {author} {\bibinfo {author} {\bibfnamefont {A.}~\bibnamefont
  {Buck}}, \bibinfo {author} {\bibfnamefont {J.}~\bibnamefont {Wenz}}, \bibinfo
  {author} {\bibfnamefont {J.}~\bibnamefont {Xu}}, \bibinfo {author}
  {\bibfnamefont {K.}~\bibnamefont {Khrennikov}}, \bibinfo {author}
  {\bibfnamefont {K.}~\bibnamefont {Schmid}}, \bibinfo {author} {\bibfnamefont
  {M.}~\bibnamefont {Heigoldt}}, \bibinfo {author} {\bibfnamefont {J.~M.}\
  \bibnamefont {Mikhailova}}, \bibinfo {author} {\bibfnamefont
  {M.}~\bibnamefont {Geissler}}, \bibinfo {author} {\bibfnamefont
  {B.}~\bibnamefont {Shen}}, \bibinfo {author} {\bibfnamefont {F.}~\bibnamefont
  {Krausz}}, \bibinfo {author} {\bibfnamefont {S.}~\bibnamefont {Karsch}},\
  and\ \bibinfo {author} {\bibfnamefont {L.}~\bibnamefont {Veisz}},\ }\bibfield
   {title} {\enquote {\bibinfo {title} {Shock-front injector for high-quality
  laser-plasma acceleration},}\ }\href
  {https://doi.org/10.1103/PhysRevLett.110.185006} {\bibfield  {journal}
  {\bibinfo  {journal} {Phys. Rev. Lett.}\ }\textbf {\bibinfo {volume} {110}},\
  \bibinfo {pages} {185006} (\bibinfo {year} {2013}{\natexlab{a}})}\BibitemShut
  {NoStop}%
\bibitem [{\citenamefont {Couperus Cabada\ifmmode~\breve{g}\else \u{g}\fi{}}\
  \emph {et~al.}(2021)\citenamefont {Couperus Cabada\ifmmode~\breve{g}\else
  \u{g}\fi{}}, \citenamefont {Pausch}, \citenamefont {Sch\"obel}, \citenamefont
  {Bussmann}, \citenamefont {Chang}, \citenamefont {Corde}, \citenamefont
  {Debus}, \citenamefont {Ding}, \citenamefont {D\"opp}, \citenamefont
  {Foerster}, \citenamefont {Gilljohann}, \citenamefont {Haberstroh},
  \citenamefont {Heinemann}, \citenamefont {Hidding}, \citenamefont {Karsch},
  \citenamefont {Koehler}, \citenamefont {Kononenko}, \citenamefont {Knetsch},
  \citenamefont {Kurz}, \citenamefont {Martinez de~la Ossa}, \citenamefont
  {Nutter}, \citenamefont {Raj}, \citenamefont {Steiniger}, \citenamefont
  {Schramm}, \citenamefont {Ufer},\ and\ \citenamefont
  {Irman}}]{Couperus2021Hybrid}%
  \BibitemOpen
  \bibfield  {author} {\bibinfo {author} {\bibfnamefont {J.~P.}\ \bibnamefont
  {Couperus Cabada\ifmmode~\breve{g}\else \u{g}\fi{}}}, \bibinfo {author}
  {\bibfnamefont {R.}~\bibnamefont {Pausch}}, \bibinfo {author} {\bibfnamefont
  {S.}~\bibnamefont {Sch\"obel}}, \bibinfo {author} {\bibfnamefont
  {M.}~\bibnamefont {Bussmann}}, \bibinfo {author} {\bibfnamefont {Y.-Y.}\
  \bibnamefont {Chang}}, \bibinfo {author} {\bibfnamefont {S.}~\bibnamefont
  {Corde}}, \bibinfo {author} {\bibfnamefont {A.}~\bibnamefont {Debus}},
  \bibinfo {author} {\bibfnamefont {H.}~\bibnamefont {Ding}}, \bibinfo {author}
  {\bibfnamefont {A.}~\bibnamefont {D\"opp}}, \bibinfo {author} {\bibfnamefont
  {F.~M.}\ \bibnamefont {Foerster}}, \bibinfo {author} {\bibfnamefont
  {M.}~\bibnamefont {Gilljohann}}, \bibinfo {author} {\bibfnamefont
  {F.}~\bibnamefont {Haberstroh}}, \bibinfo {author} {\bibfnamefont
  {T.}~\bibnamefont {Heinemann}}, \bibinfo {author} {\bibfnamefont
  {B.}~\bibnamefont {Hidding}}, \bibinfo {author} {\bibfnamefont
  {S.}~\bibnamefont {Karsch}}, \bibinfo {author} {\bibfnamefont
  {A.}~\bibnamefont {Koehler}}, \bibinfo {author} {\bibfnamefont
  {O.}~\bibnamefont {Kononenko}}, \bibinfo {author} {\bibfnamefont
  {A.}~\bibnamefont {Knetsch}}, \bibinfo {author} {\bibfnamefont
  {T.}~\bibnamefont {Kurz}}, \bibinfo {author} {\bibfnamefont {A.}~\bibnamefont
  {Martinez de~la Ossa}}, \bibinfo {author} {\bibfnamefont {A.}~\bibnamefont
  {Nutter}}, \bibinfo {author} {\bibfnamefont {G.}~\bibnamefont {Raj}},
  \bibinfo {author} {\bibfnamefont {K.}~\bibnamefont {Steiniger}}, \bibinfo
  {author} {\bibfnamefont {U.}~\bibnamefont {Schramm}}, \bibinfo {author}
  {\bibfnamefont {P.}~\bibnamefont {Ufer}},\ and\ \bibinfo {author}
  {\bibfnamefont {A.}~\bibnamefont {Irman}},\ }\bibfield  {title} {\enquote
  {\bibinfo {title} {Gas-dynamic density downramp injection in a beam-driven
  plasma wakefield accelerator},}\ }\href
  {https://doi.org/10.1103/PhysRevResearch.3.L042005} {\bibfield  {journal}
  {\bibinfo  {journal} {Phys. Rev. Research}\ }\textbf {\bibinfo {volume}
  {3}},\ \bibinfo {pages} {L042005} (\bibinfo {year} {2021})}\BibitemShut
  {NoStop}%
\bibitem [{\citenamefont {Wang}\ \emph {et~al.}(2021)\citenamefont {Wang},
  \citenamefont {Feng}, \citenamefont {Ke}, \citenamefont {Yu}, \citenamefont
  {Xu}, \citenamefont {Qi}, \citenamefont {Chen}, \citenamefont {Qin},
  \citenamefont {Zhang}, \citenamefont {Fang}, \citenamefont {Liu},
  \citenamefont {Jiang}, \citenamefont {Wang}, \citenamefont {Wang},
  \citenamefont {Yang}, \citenamefont {Wu}, \citenamefont {Leng}, \citenamefont
  {Liu}, \citenamefont {Li},\ and\ \citenamefont {Xu}}]{Wang2021FEL}%
  \BibitemOpen
  \bibfield  {author} {\bibinfo {author} {\bibfnamefont {W.}~\bibnamefont
  {Wang}}, \bibinfo {author} {\bibfnamefont {K.}~\bibnamefont {Feng}}, \bibinfo
  {author} {\bibfnamefont {L.}~\bibnamefont {Ke}}, \bibinfo {author}
  {\bibfnamefont {C.}~\bibnamefont {Yu}}, \bibinfo {author} {\bibfnamefont
  {Y.}~\bibnamefont {Xu}}, \bibinfo {author} {\bibfnamefont {R.}~\bibnamefont
  {Qi}}, \bibinfo {author} {\bibfnamefont {Y.}~\bibnamefont {Chen}}, \bibinfo
  {author} {\bibfnamefont {Z.}~\bibnamefont {Qin}}, \bibinfo {author}
  {\bibfnamefont {Z.}~\bibnamefont {Zhang}}, \bibinfo {author} {\bibfnamefont
  {M.}~\bibnamefont {Fang}}, \bibinfo {author} {\bibfnamefont {J.}~\bibnamefont
  {Liu}}, \bibinfo {author} {\bibfnamefont {K.}~\bibnamefont {Jiang}}, \bibinfo
  {author} {\bibfnamefont {H.}~\bibnamefont {Wang}}, \bibinfo {author}
  {\bibfnamefont {C.}~\bibnamefont {Wang}}, \bibinfo {author} {\bibfnamefont
  {X.}~\bibnamefont {Yang}}, \bibinfo {author} {\bibfnamefont {F.}~\bibnamefont
  {Wu}}, \bibinfo {author} {\bibfnamefont {Y.}~\bibnamefont {Leng}}, \bibinfo
  {author} {\bibfnamefont {J.}~\bibnamefont {Liu}}, \bibinfo {author}
  {\bibfnamefont {R.}~\bibnamefont {Li}},\ and\ \bibinfo {author}
  {\bibfnamefont {Z.}~\bibnamefont {Xu}},\ }\bibfield  {title} {\enquote
  {\bibinfo {title} {Free-electron lasing at 27 nanometres based on a laser
  wakefield accelerator},}\ }\href {https://doi.org/10.1038/s41586-021-03678-x}
  {\bibfield  {journal} {\bibinfo  {journal} {Nature}\ }\textbf {\bibinfo
  {volume} {595}},\ \bibinfo {pages} {516--520} (\bibinfo {year}
  {2021})}\BibitemShut {NoStop}%
\bibitem [{\citenamefont {Yeboah}, \citenamefont {Sandison},\ and\
  \citenamefont {Moskvin}(2002)}]{Yeboah_2002_250MeVEleTherapy}%
  \BibitemOpen
  \bibfield  {author} {\bibinfo {author} {\bibfnamefont {C.}~\bibnamefont
  {Yeboah}}, \bibinfo {author} {\bibfnamefont {G.~A.}\ \bibnamefont
  {Sandison}},\ and\ \bibinfo {author} {\bibfnamefont {V.}~\bibnamefont
  {Moskvin}},\ }\bibfield  {title} {\enquote {\bibinfo {title} {Optimization of
  intensity-modulated very high energy (50{\textendash}250 {MeV}) electron
  therapy},}\ }\href {https://doi.org/10.1088/0031-9155/47/8/305} {\bibfield
  {journal} {\bibinfo  {journal} {Physics in Medicine and Biology}\ }\textbf
  {\bibinfo {volume} {47}},\ \bibinfo {pages} {1285--1301} (\bibinfo {year}
  {2002})}\BibitemShut {NoStop}%
\bibitem [{\citenamefont {Glinec}\ \emph
  {et~al.}(2006{\natexlab{a}})\citenamefont {Glinec}, \citenamefont {Faure},
  \citenamefont {Malka}, \citenamefont {Fuchs}, \citenamefont {Szymanowski},\
  and\ \citenamefont {Oelfke}}]{Glinnec2005}%
  \BibitemOpen
  \bibfield  {author} {\bibinfo {author} {\bibfnamefont {Y.}~\bibnamefont
  {Glinec}}, \bibinfo {author} {\bibfnamefont {J.}~\bibnamefont {Faure}},
  \bibinfo {author} {\bibfnamefont {V.}~\bibnamefont {Malka}}, \bibinfo
  {author} {\bibfnamefont {T.}~\bibnamefont {Fuchs}}, \bibinfo {author}
  {\bibfnamefont {H.}~\bibnamefont {Szymanowski}},\ and\ \bibinfo {author}
  {\bibfnamefont {U.}~\bibnamefont {Oelfke}},\ }\bibfield  {title} {\enquote
  {\bibinfo {title} {Radiotherapy with laser-plasma accelerators: Monte carlo
  simulation of dose deposited by an experimental quasimonoenergetic electron
  beam},}\ }\href {https://doi.org/https://doi.org/10.1118/1.2140115}
  {\bibfield  {journal} {\bibinfo  {journal} {Medical Physics}\ }\textbf
  {\bibinfo {volume} {33}},\ \bibinfo {pages} {155--162} (\bibinfo {year}
  {2006}{\natexlab{a}})},\ \Eprint
  {https://arxiv.org/abs/https://aapm.onlinelibrary.wiley.com/doi/pdf/10.1118/1.2140115}
  {https://aapm.onlinelibrary.wiley.com/doi/pdf/10.1118/1.2140115} \BibitemShut
  {NoStop}%
\bibitem [{\citenamefont {Malka}\ \emph {et~al.}(2008)\citenamefont {Malka},
  \citenamefont {Faure}, \citenamefont {Gauduel}, \citenamefont {Lefebvre},
  \citenamefont {Rousse},\ and\ \citenamefont {Phuoc}}]{Malka2008}%
  \BibitemOpen
  \bibfield  {author} {\bibinfo {author} {\bibfnamefont {V.}~\bibnamefont
  {Malka}}, \bibinfo {author} {\bibfnamefont {J.}~\bibnamefont {Faure}},
  \bibinfo {author} {\bibfnamefont {Y.~A.}\ \bibnamefont {Gauduel}}, \bibinfo
  {author} {\bibfnamefont {E.}~\bibnamefont {Lefebvre}}, \bibinfo {author}
  {\bibfnamefont {A.}~\bibnamefont {Rousse}},\ and\ \bibinfo {author}
  {\bibfnamefont {K.~T.}\ \bibnamefont {Phuoc}},\ }\bibfield  {title} {\enquote
  {\bibinfo {title} {Principles and applications of compact laser–plasma
  accelerators},}\ }\href {https://doi.org/10.1038/nphys966} {\bibfield
  {journal} {\bibinfo  {journal} {Nature Physics}\ }\textbf {\bibinfo {volume}
  {4}},\ \bibinfo {pages} {447–453} (\bibinfo {year} {2008})},\ \Eprint
  {https://arxiv.org/abs/https://doi.org/110.1038/nphys966}
  {https://doi.org/110.1038/nphys966} \BibitemShut {NoStop}%
\bibitem [{\citenamefont {Fuchs}\ \emph {et~al.}(2009)\citenamefont {Fuchs},
  \citenamefont {Szymanowski}, \citenamefont {Oelfke}, \citenamefont {Glinec},
  \citenamefont {Rechatin}, \citenamefont {Faure},\ and\ \citenamefont
  {Malka}}]{Fuchs_2009}%
  \BibitemOpen
  \bibfield  {author} {\bibinfo {author} {\bibfnamefont {T.}~\bibnamefont
  {Fuchs}}, \bibinfo {author} {\bibfnamefont {H.}~\bibnamefont {Szymanowski}},
  \bibinfo {author} {\bibfnamefont {U.}~\bibnamefont {Oelfke}}, \bibinfo
  {author} {\bibfnamefont {Y.}~\bibnamefont {Glinec}}, \bibinfo {author}
  {\bibfnamefont {C.}~\bibnamefont {Rechatin}}, \bibinfo {author}
  {\bibfnamefont {J.}~\bibnamefont {Faure}},\ and\ \bibinfo {author}
  {\bibfnamefont {V.}~\bibnamefont {Malka}},\ }\bibfield  {title} {\enquote
  {\bibinfo {title} {Treatment planning for laser-accelerated very-high energy
  electrons},}\ }\href {https://doi.org/10.1088/0031-9155/54/11/003} {\bibfield
   {journal} {\bibinfo  {journal} {Physics in Medicine and Biology}\ }\textbf
  {\bibinfo {volume} {54}},\ \bibinfo {pages} {3315--3328} (\bibinfo {year}
  {2009})}\BibitemShut {NoStop}%
\bibitem [{\citenamefont {Massimo}\ \emph {et~al.}(2018)\citenamefont
  {Massimo}, \citenamefont {Lifschitz}, \citenamefont {Thaury},\ and\
  \citenamefont {Malka}}]{Massimo_2018}%
  \BibitemOpen
  \bibfield  {author} {\bibinfo {author} {\bibfnamefont {F.}~\bibnamefont
  {Massimo}}, \bibinfo {author} {\bibfnamefont {A.~F.}\ \bibnamefont
  {Lifschitz}}, \bibinfo {author} {\bibfnamefont {C.}~\bibnamefont {Thaury}},\
  and\ \bibinfo {author} {\bibfnamefont {V.}~\bibnamefont {Malka}},\ }\bibfield
   {title} {\enquote {\bibinfo {title} {Numerical study of laser energy effects
  on density transition injection in laser wakefield acceleration},}\ }\href
  {https://doi.org/10.1088/1361-6587/aaa336} {\bibfield  {journal} {\bibinfo
  {journal} {Plasma Physics and Controlled Fusion}\ }\textbf {\bibinfo {volume}
  {60}},\ \bibinfo {pages} {034005} (\bibinfo {year} {2018})}\BibitemShut
  {NoStop}%
\bibitem [{\citenamefont {Massimo}\ \emph {et~al.}(2017)\citenamefont
  {Massimo}, \citenamefont {Lifschitz}, \citenamefont {Thaury},\ and\
  \citenamefont {Malka}}]{Massimo_2017}%
  \BibitemOpen
  \bibfield  {author} {\bibinfo {author} {\bibfnamefont {F.}~\bibnamefont
  {Massimo}}, \bibinfo {author} {\bibfnamefont {A.~F.}\ \bibnamefont
  {Lifschitz}}, \bibinfo {author} {\bibfnamefont {C.}~\bibnamefont {Thaury}},\
  and\ \bibinfo {author} {\bibfnamefont {V.}~\bibnamefont {Malka}},\ }\bibfield
   {title} {\enquote {\bibinfo {title} {Numerical studies of density transition
  injection in laser wakefield acceleration},}\ }\href
  {https://doi.org/10.1088/1361-6587/aa717d} {\bibfield  {journal} {\bibinfo
  {journal} {Plasma Physics and Controlled Fusion}\ }\textbf {\bibinfo {volume}
  {59}},\ \bibinfo {pages} {085004} (\bibinfo {year} {2017})}\BibitemShut
  {NoStop}%
\bibitem [{\citenamefont {Lehe}\ \emph {et~al.}(2016)\citenamefont {Lehe},
  \citenamefont {Kirchen}, \citenamefont {Andriyash}, \citenamefont {Godfrey},\
  and\ \citenamefont {Vay}}]{LEHE201666}%
  \BibitemOpen
  \bibfield  {author} {\bibinfo {author} {\bibfnamefont {R.}~\bibnamefont
  {Lehe}}, \bibinfo {author} {\bibfnamefont {M.}~\bibnamefont {Kirchen}},
  \bibinfo {author} {\bibfnamefont {I.~A.}\ \bibnamefont {Andriyash}}, \bibinfo
  {author} {\bibfnamefont {B.~B.}\ \bibnamefont {Godfrey}},\ and\ \bibinfo
  {author} {\bibfnamefont {J.-L.}\ \bibnamefont {Vay}},\ }\bibfield  {title}
  {\enquote {\bibinfo {title} {A spectral, quasi-cylindrical and
  dispersion-free particle-in-cell algorithm},}\ }\href
  {https://doi.org/https://doi.org/10.1016/j.cpc.2016.02.007} {\bibfield
  {journal} {\bibinfo  {journal} {Computer Physics Communications}\ }\textbf
  {\bibinfo {volume} {203}},\ \bibinfo {pages} {66--82} (\bibinfo {year}
  {2016})}\BibitemShut {NoStop}%
\bibitem [{\citenamefont {Kroupp}\ \emph {et~al.}(2022)\citenamefont {Kroupp},
  \citenamefont {Tata}, \citenamefont {Wan}, \citenamefont {Levy},
  \citenamefont {Smartsev}, \citenamefont {Levine}, \citenamefont {Seemann},
  \citenamefont {Adelberg}, \citenamefont {Piliposian}, \citenamefont
  {Queller}, \citenamefont {Segre}, \citenamefont {Phuoc}, \citenamefont
  {Kozlova},\ and\ \citenamefont {Malka}}]{Kroupp2022}%
  \BibitemOpen
  \bibfield  {author} {\bibinfo {author} {\bibfnamefont {E.}~\bibnamefont
  {Kroupp}}, \bibinfo {author} {\bibfnamefont {S.}~\bibnamefont {Tata}},
  \bibinfo {author} {\bibfnamefont {Y.}~\bibnamefont {Wan}}, \bibinfo {author}
  {\bibfnamefont {D.}~\bibnamefont {Levy}}, \bibinfo {author} {\bibfnamefont
  {S.}~\bibnamefont {Smartsev}}, \bibinfo {author} {\bibfnamefont
  {E.}~\bibnamefont {Levine}}, \bibinfo {author} {\bibfnamefont
  {O.}~\bibnamefont {Seemann}}, \bibinfo {author} {\bibfnamefont
  {M.}~\bibnamefont {Adelberg}}, \bibinfo {author} {\bibfnamefont
  {R.}~\bibnamefont {Piliposian}}, \bibinfo {author} {\bibfnamefont
  {T.}~\bibnamefont {Queller}}, \bibinfo {author} {\bibfnamefont
  {E.}~\bibnamefont {Segre}}, \bibinfo {author} {\bibfnamefont
  {K.}~\bibnamefont {Phuoc}}, \bibinfo {author} {\bibfnamefont
  {M.}~\bibnamefont {Kozlova}},\ and\ \bibinfo {author} {\bibfnamefont
  {V.}~\bibnamefont {Malka}},\ }\bibfield  {title} {\enquote {\bibinfo {title}
  {Commissioning and first results from the new 2 × 100 tw laser at the
  wis},}\ }\href {https://doi.org/10.1063/5.0090514} {\bibfield  {journal}
  {\bibinfo  {journal} {Matter and Radiation at Extremes}\ }\textbf {\bibinfo
  {volume} {7}},\ \bibinfo {pages} {044401} (\bibinfo {year}
  {2022})}\BibitemShut {NoStop}%
\bibitem [{\citenamefont {Buck}\ \emph
  {et~al.}(2013{\natexlab{b}})\citenamefont {Buck}, \citenamefont {Wenz},
  \citenamefont {Xu}, \citenamefont {Khrennikov}, \citenamefont {Schmid},
  \citenamefont {Heigoldt}, \citenamefont {Mikhailova}, \citenamefont
  {Geissler}, \citenamefont {Shen}, \citenamefont {Krausz}, \citenamefont
  {Karsch},\ and\ \citenamefont {Veisz}}]{BUCK2013}%
  \BibitemOpen
  \bibfield  {author} {\bibinfo {author} {\bibfnamefont {A.}~\bibnamefont
  {Buck}}, \bibinfo {author} {\bibfnamefont {J.}~\bibnamefont {Wenz}}, \bibinfo
  {author} {\bibfnamefont {J.}~\bibnamefont {Xu}}, \bibinfo {author}
  {\bibfnamefont {K.}~\bibnamefont {Khrennikov}}, \bibinfo {author}
  {\bibfnamefont {K.}~\bibnamefont {Schmid}}, \bibinfo {author} {\bibfnamefont
  {M.}~\bibnamefont {Heigoldt}}, \bibinfo {author} {\bibfnamefont {J.~M.}\
  \bibnamefont {Mikhailova}}, \bibinfo {author} {\bibfnamefont
  {M.}~\bibnamefont {Geissler}}, \bibinfo {author} {\bibfnamefont
  {B.}~\bibnamefont {Shen}}, \bibinfo {author} {\bibfnamefont {F.}~\bibnamefont
  {Krausz}}, \bibinfo {author} {\bibfnamefont {S.}~\bibnamefont {Karsch}},\
  and\ \bibinfo {author} {\bibfnamefont {L.}~\bibnamefont {Veisz}},\ }\bibfield
   {title} {\enquote {\bibinfo {title} {Shock-front injector for high-quality
  laser-plasma acceleration},}\ }\href
  {https://doi.org/10.1103/PhysRevLett.110.185006} {\bibfield  {journal}
  {\bibinfo  {journal} {Phys. Rev. Lett.}\ }\textbf {\bibinfo {volume} {110}},\
  \bibinfo {pages} {185006} (\bibinfo {year} {2013}{\natexlab{b}})}\BibitemShut
  {NoStop}%
\bibitem [{\citenamefont {Schmid}\ \emph {et~al.}(2010)\citenamefont {Schmid},
  \citenamefont {Buck}, \citenamefont {Sears}, \citenamefont {Mikhailova},
  \citenamefont {Tautz}, \citenamefont {Herrmann}, \citenamefont {Geissler},
  \citenamefont {Krausz},\ and\ \citenamefont {Veisz}}]{Schmid2010}%
  \BibitemOpen
  \bibfield  {author} {\bibinfo {author} {\bibfnamefont {K.}~\bibnamefont
  {Schmid}}, \bibinfo {author} {\bibfnamefont {A.}~\bibnamefont {Buck}},
  \bibinfo {author} {\bibfnamefont {C.~M.~S.}\ \bibnamefont {Sears}}, \bibinfo
  {author} {\bibfnamefont {J.~M.}\ \bibnamefont {Mikhailova}}, \bibinfo
  {author} {\bibfnamefont {R.}~\bibnamefont {Tautz}}, \bibinfo {author}
  {\bibfnamefont {D.}~\bibnamefont {Herrmann}}, \bibinfo {author}
  {\bibfnamefont {M.}~\bibnamefont {Geissler}}, \bibinfo {author}
  {\bibfnamefont {F.}~\bibnamefont {Krausz}},\ and\ \bibinfo {author}
  {\bibfnamefont {L.}~\bibnamefont {Veisz}},\ }\bibfield  {title} {\enquote
  {\bibinfo {title} {Density-transition based electron injector for laser
  driven wakefield accelerators},}\ }\href
  {https://doi.org/10.1103/PhysRevSTAB.13.091301} {\bibfield  {journal}
  {\bibinfo  {journal} {Phys. Rev. ST Accel. Beams}\ }\textbf {\bibinfo
  {volume} {13}},\ \bibinfo {pages} {091301} (\bibinfo {year}
  {2010})}\BibitemShut {NoStop}%
\bibitem [{\citenamefont {Sun}\ \emph {et~al.}(1987)\citenamefont {Sun},
  \citenamefont {Ott}, \citenamefont {Lee},\ and\ \citenamefont
  {Guzdar}}]{Sum1987_selffocusing}%
  \BibitemOpen
  \bibfield  {author} {\bibinfo {author} {\bibfnamefont {G.}~\bibnamefont
  {Sun}}, \bibinfo {author} {\bibfnamefont {E.}~\bibnamefont {Ott}}, \bibinfo
  {author} {\bibfnamefont {Y.~C.}\ \bibnamefont {Lee}},\ and\ \bibinfo {author}
  {\bibfnamefont {P.}~\bibnamefont {Guzdar}},\ }\bibfield  {title} {\enquote
  {\bibinfo {title} {Self‐focusing of short intense pulses in plasmas},}\
  }\href {https://doi.org/10.1063/1.866349} {\bibfield  {journal} {\bibinfo
  {journal} {The Physics of Fluids}\ }\textbf {\bibinfo {volume} {30}},\
  \bibinfo {pages} {526--532} (\bibinfo {year} {1987})},\ \Eprint
  {https://arxiv.org/abs/https://aip.scitation.org/doi/pdf/10.1063/1.866349}
  {https://aip.scitation.org/doi/pdf/10.1063/1.866349} \BibitemShut {NoStop}%
\bibitem [{\citenamefont {Vieira}\ \emph {et~al.}(2010)\citenamefont {Vieira},
  \citenamefont {Fi{\'{u}}za}, \citenamefont {Silva}, \citenamefont
  {Tzoufras},\ and\ \citenamefont {Mori}}]{Vieira_2010}%
  \BibitemOpen
  \bibfield  {author} {\bibinfo {author} {\bibfnamefont {J.}~\bibnamefont
  {Vieira}}, \bibinfo {author} {\bibfnamefont {F.}~\bibnamefont {Fi{\'{u}}za}},
  \bibinfo {author} {\bibfnamefont {L.~O.}\ \bibnamefont {Silva}}, \bibinfo
  {author} {\bibfnamefont {M.}~\bibnamefont {Tzoufras}},\ and\ \bibinfo
  {author} {\bibfnamefont {W.~B.}\ \bibnamefont {Mori}},\ }\bibfield  {title}
  {\enquote {\bibinfo {title} {Onset of self-steepening of intense laser pulses
  in plasmas},}\ }\href {https://doi.org/10.1088/1367-2630/12/4/045025}
  {\bibfield  {journal} {\bibinfo  {journal} {New Journal of Physics}\ }\textbf
  {\bibinfo {volume} {12}},\ \bibinfo {pages} {045025} (\bibinfo {year}
  {2010})}\BibitemShut {NoStop}%
\bibitem [{\citenamefont {Xu}\ \emph {et~al.}(2017)\citenamefont {Xu},
  \citenamefont {Li}, \citenamefont {An}, \citenamefont {Dalichaouch},
  \citenamefont {Yu}, \citenamefont {Lu}, \citenamefont {Joshi},\ and\
  \citenamefont {Mori}}]{XU2017}%
  \BibitemOpen
  \bibfield  {author} {\bibinfo {author} {\bibfnamefont {X.~L.}\ \bibnamefont
  {Xu}}, \bibinfo {author} {\bibfnamefont {F.}~\bibnamefont {Li}}, \bibinfo
  {author} {\bibfnamefont {W.}~\bibnamefont {An}}, \bibinfo {author}
  {\bibfnamefont {T.~N.}\ \bibnamefont {Dalichaouch}}, \bibinfo {author}
  {\bibfnamefont {P.}~\bibnamefont {Yu}}, \bibinfo {author} {\bibfnamefont
  {W.}~\bibnamefont {Lu}}, \bibinfo {author} {\bibfnamefont {C.}~\bibnamefont
  {Joshi}},\ and\ \bibinfo {author} {\bibfnamefont {W.~B.}\ \bibnamefont
  {Mori}},\ }\bibfield  {title} {\enquote {\bibinfo {title} {High quality
  electron bunch generation using a longitudinal density-tailored plasma-based
  accelerator in the three-dimensional blowout regime},}\ }\href
  {https://doi.org/10.1103/PhysRevAccelBeams.20.111303} {\bibfield  {journal}
  {\bibinfo  {journal} {Phys. Rev. Accel. Beams}\ }\textbf {\bibinfo {volume}
  {20}},\ \bibinfo {pages} {111303} (\bibinfo {year} {2017})}\BibitemShut
  {NoStop}%
\bibitem [{\citenamefont {Tzoufras}\ \emph {et~al.}(2008)\citenamefont
  {Tzoufras}, \citenamefont {Lu}, \citenamefont {Tsung}, \citenamefont {Huang},
  \citenamefont {Mori}, \citenamefont {Katsouleas}, \citenamefont {Vieira},
  \citenamefont {Fonseca},\ and\ \citenamefont {Silva}}]{Tzoufras2008}%
  \BibitemOpen
  \bibfield  {author} {\bibinfo {author} {\bibfnamefont {M.}~\bibnamefont
  {Tzoufras}}, \bibinfo {author} {\bibfnamefont {W.}~\bibnamefont {Lu}},
  \bibinfo {author} {\bibfnamefont {F.~S.}\ \bibnamefont {Tsung}}, \bibinfo
  {author} {\bibfnamefont {C.}~\bibnamefont {Huang}}, \bibinfo {author}
  {\bibfnamefont {W.~B.}\ \bibnamefont {Mori}}, \bibinfo {author}
  {\bibfnamefont {T.}~\bibnamefont {Katsouleas}}, \bibinfo {author}
  {\bibfnamefont {J.}~\bibnamefont {Vieira}}, \bibinfo {author} {\bibfnamefont
  {R.~A.}\ \bibnamefont {Fonseca}},\ and\ \bibinfo {author} {\bibfnamefont
  {L.~O.}\ \bibnamefont {Silva}},\ }\bibfield  {title} {\enquote {\bibinfo
  {title} {Beam loading in the nonlinear regime of plasma-based
  acceleration},}\ }\href {https://doi.org/10.1103/PhysRevLett.101.145002}
  {\bibfield  {journal} {\bibinfo  {journal} {Phys. Rev. Lett.}\ }\textbf
  {\bibinfo {volume} {101}},\ \bibinfo {pages} {145002} (\bibinfo {year}
  {2008})}\BibitemShut {NoStop}%
\bibitem [{\citenamefont {Lu}\ \emph {et~al.}(2007)\citenamefont {Lu},
  \citenamefont {Tzoufras}, \citenamefont {Joshi}, \citenamefont {Tsung},
  \citenamefont {Mori}, \citenamefont {Vieira}, \citenamefont {Fonseca},\ and\
  \citenamefont {Silva}}]{LU2007LaserSelfGuiding}%
  \BibitemOpen
  \bibfield  {author} {\bibinfo {author} {\bibfnamefont {W.}~\bibnamefont
  {Lu}}, \bibinfo {author} {\bibfnamefont {M.}~\bibnamefont {Tzoufras}},
  \bibinfo {author} {\bibfnamefont {C.}~\bibnamefont {Joshi}}, \bibinfo
  {author} {\bibfnamefont {F.~S.}\ \bibnamefont {Tsung}}, \bibinfo {author}
  {\bibfnamefont {W.~B.}\ \bibnamefont {Mori}}, \bibinfo {author}
  {\bibfnamefont {J.}~\bibnamefont {Vieira}}, \bibinfo {author} {\bibfnamefont
  {R.~A.}\ \bibnamefont {Fonseca}},\ and\ \bibinfo {author} {\bibfnamefont
  {L.~O.}\ \bibnamefont {Silva}},\ }\bibfield  {title} {\enquote {\bibinfo
  {title} {Generating multi-gev electron bunches using single stage laser
  wakefield acceleration in a 3d nonlinear regime},}\ }\href
  {https://doi.org/10.1103/PhysRevSTAB.10.061301} {\bibfield  {journal}
  {\bibinfo  {journal} {Phys. Rev. ST Accel. Beams}\ }\textbf {\bibinfo
  {volume} {10}},\ \bibinfo {pages} {061301} (\bibinfo {year}
  {2007})}\BibitemShut {NoStop}%
\bibitem [{\citenamefont {G\"otzfried}\ \emph {et~al.}(2020)\citenamefont
  {G\"otzfried}, \citenamefont {D\"opp}, \citenamefont {Gilljohann},
  \citenamefont {Foerster}, \citenamefont {Ding}, \citenamefont {Schindler},
  \citenamefont {Schilling}, \citenamefont {Buck}, \citenamefont {Veisz},\ and\
  \citenamefont {Karsch}}]{Gotzfried2020}%
  \BibitemOpen
  \bibfield  {author} {\bibinfo {author} {\bibfnamefont {J.}~\bibnamefont
  {G\"otzfried}}, \bibinfo {author} {\bibfnamefont {A.}~\bibnamefont {D\"opp}},
  \bibinfo {author} {\bibfnamefont {M.~F.}\ \bibnamefont {Gilljohann}},
  \bibinfo {author} {\bibfnamefont {F.~M.}\ \bibnamefont {Foerster}}, \bibinfo
  {author} {\bibfnamefont {H.}~\bibnamefont {Ding}}, \bibinfo {author}
  {\bibfnamefont {S.}~\bibnamefont {Schindler}}, \bibinfo {author}
  {\bibfnamefont {G.}~\bibnamefont {Schilling}}, \bibinfo {author}
  {\bibfnamefont {A.}~\bibnamefont {Buck}}, \bibinfo {author} {\bibfnamefont
  {L.}~\bibnamefont {Veisz}},\ and\ \bibinfo {author} {\bibfnamefont
  {S.}~\bibnamefont {Karsch}},\ }\bibfield  {title} {\enquote {\bibinfo {title}
  {Physics of high-charge electron beams in laser-plasma wakefields},}\ }\href
  {https://doi.org/10.1103/PhysRevX.10.041015} {\bibfield  {journal} {\bibinfo
  {journal} {Phys. Rev. X}\ }\textbf {\bibinfo {volume} {10}},\ \bibinfo
  {pages} {041015} (\bibinfo {year} {2020})}\BibitemShut {NoStop}%
\bibitem [{\citenamefont {Corde}\ \emph {et~al.}(2015)\citenamefont {Corde},
  \citenamefont {Adli}, \citenamefont {Allen}, \citenamefont {An},
  \citenamefont {Clarke}, \citenamefont {Clayton}, \citenamefont {Delahaye},
  \citenamefont {Frederico}, \citenamefont {Gessner}, \citenamefont {Green},
  \citenamefont {Hogan}, \citenamefont {Joshi}, \citenamefont {Lipkowitz},
  \citenamefont {Litos}, \citenamefont {Lu}, \citenamefont {Marsh},
  \citenamefont {Mori}, \citenamefont {Schmeltz}, \citenamefont
  {Vafaei-Najafabadi}, \citenamefont {Walz}, \citenamefont {Yakimenko},\ and\
  \citenamefont {Yocky}}]{Corde2015}%
  \BibitemOpen
  \bibfield  {author} {\bibinfo {author} {\bibfnamefont {S.}~\bibnamefont
  {Corde}}, \bibinfo {author} {\bibfnamefont {E.}~\bibnamefont {Adli}},
  \bibinfo {author} {\bibfnamefont {J.~M.}\ \bibnamefont {Allen}}, \bibinfo
  {author} {\bibfnamefont {W.}~\bibnamefont {An}}, \bibinfo {author}
  {\bibfnamefont {C.~I.}\ \bibnamefont {Clarke}}, \bibinfo {author}
  {\bibfnamefont {C.~E.}\ \bibnamefont {Clayton}}, \bibinfo {author}
  {\bibfnamefont {J.~P.}\ \bibnamefont {Delahaye}}, \bibinfo {author}
  {\bibfnamefont {J.}~\bibnamefont {Frederico}}, \bibinfo {author}
  {\bibfnamefont {S.}~\bibnamefont {Gessner}}, \bibinfo {author} {\bibfnamefont
  {S.~Z.}\ \bibnamefont {Green}}, \bibinfo {author} {\bibfnamefont {M.~J.}\
  \bibnamefont {Hogan}}, \bibinfo {author} {\bibfnamefont {C.}~\bibnamefont
  {Joshi}}, \bibinfo {author} {\bibfnamefont {N.}~\bibnamefont {Lipkowitz}},
  \bibinfo {author} {\bibfnamefont {M.}~\bibnamefont {Litos}}, \bibinfo
  {author} {\bibfnamefont {W.}~\bibnamefont {Lu}}, \bibinfo {author}
  {\bibfnamefont {K.~A.}\ \bibnamefont {Marsh}}, \bibinfo {author}
  {\bibfnamefont {W.~B.}\ \bibnamefont {Mori}}, \bibinfo {author}
  {\bibfnamefont {M.}~\bibnamefont {Schmeltz}}, \bibinfo {author}
  {\bibfnamefont {N.}~\bibnamefont {Vafaei-Najafabadi}}, \bibinfo {author}
  {\bibfnamefont {D.}~\bibnamefont {Walz}}, \bibinfo {author} {\bibfnamefont
  {V.}~\bibnamefont {Yakimenko}},\ and\ \bibinfo {author} {\bibfnamefont
  {G.}~\bibnamefont {Yocky}},\ }\bibfield  {title} {\enquote {\bibinfo {title}
  {Multi-gigaelectronvolt acceleration of positrons in a self-loaded plasma
  wakefield},}\ }\href {https://doi.org/10.1038/nature14890} {\bibfield
  {journal} {\bibinfo  {journal} {Nature}\ }\textbf {\bibinfo {volume} {524}},\
  \bibinfo {pages} {442--445} (\bibinfo {year} {2015})}\BibitemShut {NoStop}%
\bibitem [{\citenamefont {Glinec}\ \emph
  {et~al.}(2006{\natexlab{b}})\citenamefont {Glinec}, \citenamefont {Faure},
  \citenamefont {Malka}, \citenamefont {Fuchs}, \citenamefont {Szymanowski},\
  and\ \citenamefont {Oelfke}}]{Glinec2006}%
  \BibitemOpen
  \bibfield  {author} {\bibinfo {author} {\bibfnamefont {Y.}~\bibnamefont
  {Glinec}}, \bibinfo {author} {\bibfnamefont {J.}~\bibnamefont {Faure}},
  \bibinfo {author} {\bibfnamefont {V.}~\bibnamefont {Malka}}, \bibinfo
  {author} {\bibfnamefont {T.}~\bibnamefont {Fuchs}}, \bibinfo {author}
  {\bibfnamefont {H.}~\bibnamefont {Szymanowski}},\ and\ \bibinfo {author}
  {\bibfnamefont {U.}~\bibnamefont {Oelfke}},\ }\bibfield  {title} {\enquote
  {\bibinfo {title} {Radiotherapy with laser-plasma accelerators: Monte carlo
  simulation of dose deposited by an experimental quasimonoenergetic electron
  beam},}\ }\href {https://doi.org/https://doi.org/10.1118/1.2140115}
  {\bibfield  {journal} {\bibinfo  {journal} {Medical Physics}\ }\textbf
  {\bibinfo {volume} {33}},\ \bibinfo {pages} {155--162} (\bibinfo {year}
  {2006}{\natexlab{b}})},\ \Eprint
  {https://arxiv.org/abs/https://aapm.onlinelibrary.wiley.com/doi/pdf/10.1118/1.2140115}
  {https://aapm.onlinelibrary.wiley.com/doi/pdf/10.1118/1.2140115} \BibitemShut
  {NoStop}%
\bibitem [{\citenamefont {Brahme}\ and\ \citenamefont
  {Reistad}(1972)}]{VHEE1972}%
  \BibitemOpen
  \bibfield  {author} {\bibinfo {author} {\bibfnamefont {A.}~\bibnamefont
  {Brahme}}\ and\ \bibinfo {author} {\bibfnamefont {D.}~\bibnamefont
  {Reistad}},\ }\href
  {http://inis.iaea.org/search/search.aspx?orig_q=RN:04051267} {\enquote
  {\bibinfo {title} {The microtron, a new accelerator for radiation therapy},}\
  }\bibinfo {type} {Tech. Rep.}\ (\bibinfo {address} {Sweden},\ \bibinfo {year}
  {1972})\BibitemShut {NoStop}%
\bibitem [{\citenamefont {Lu}\ \emph {et~al.}(2006)\citenamefont {Lu},
  \citenamefont {Huang}, \citenamefont {Zhou}, \citenamefont {Mori},\ and\
  \citenamefont {Katsouleas}}]{Lu2006}%
  \BibitemOpen
  \bibfield  {author} {\bibinfo {author} {\bibfnamefont {W.}~\bibnamefont
  {Lu}}, \bibinfo {author} {\bibfnamefont {C.}~\bibnamefont {Huang}}, \bibinfo
  {author} {\bibfnamefont {M.}~\bibnamefont {Zhou}}, \bibinfo {author}
  {\bibfnamefont {W.~B.}\ \bibnamefont {Mori}},\ and\ \bibinfo {author}
  {\bibfnamefont {T.}~\bibnamefont {Katsouleas}},\ }\bibfield  {title}
  {\enquote {\bibinfo {title} {Nonlinear theory for relativistic plasma
  wakefields in the blowout regime},}\ }\href
  {https://doi.org/10.1103/PhysRevLett.96.165002} {\bibfield  {journal}
  {\bibinfo  {journal} {Phys. Rev. Lett.}\ }\textbf {\bibinfo {volume} {96}},\
  \bibinfo {pages} {165002} (\bibinfo {year} {2006})}\BibitemShut {NoStop}%
\end{thebibliography}%

\end{document}